\documentclass[aps,prc,nofootinbib]{revtex4-2}
\usepackage{amsmath,amssymb}
\usepackage{graphicx}

\newcommand{\be}{\begin{equation}}
\newcommand{\ee}{\end{equation}}
\newcommand{\ba}{\begin{eqnarray}}
\newcommand{\ea}{\end{eqnarray}}
\newcommand{\ban}{\begin{eqnarray*}}
\newcommand{\ean}{\end{eqnarray*}}
\newcommand \nn {\nonumber}

\begin{document}

\title{Stability of Classical Chromodynamic Fields}

\author{Sylwia Bazak$^1$\footnote{e-mail: sylwia.bazak@gmail.com} and Stanis\l aw Mr\' owczy\' nski$^{1,2}$\footnote{e-mail: stanislaw.mrowczynski@ncbj.gov.pl}}

\affiliation{$^1$Institute of Physics, Jan Kochanowski University, ul. Uniwersytecka 7, PL-25-406 Kielce, Poland 
\\
$^2$National Centre for Nuclear Research, ul. Pasteura 7,  PL-02-093 Warsaw, Poland}

\date{February 11, 2022}

\begin{abstract}

A system of gluon fields generated at the earliest phase of relativistic heavy-ion collisions can be described in terms of classical fields. Numerical simulations show that the system is unstable but a character of the instability is not well understood. With the intention to systematically study the problem, we analyze a stability of classical chromomagnetic and chromoelectric fields which are constant and uniform. We consider the Abelian configurations discussed in the past where the fields are due to the single-color potentials linearly depending on coordinates. However, we mostly focus on the nonAbelian configurations where the fields are generated by the multi-color non-commuting constant uniform potentials. We derive a complete spectrum of small fluctuations around the background fields which obey the linearized Yang-Mills equations. The spectra of Abelian and nonAbelian configurations are similar but different and they both include unstable modes. We briefly discuss the relevance of our results for fields which are uniform only in a limited spatial domain.

\end{abstract}

\maketitle

\section{Introduction}

Soon after the discovery of asymptotic freedom \cite{Gross:1973id,Politzer:1973fx}, when quantum chromodynamics was recognized as an underlying theory of strong interactions, the stability of various configurations of classical chromodynamic fields was investigated \cite{Mandula:1976uh,Mandula:1976xf,Chang:1979tg,Sikivie:1979bq,Tudron:1980gq,Passarino:1986gs}. These studies, which revealed that numerous configurations are actually unstable, were not performed with a specific application in mind, rather it was about better understanding the newborn theory.

Today, classical chromodynamics is often used as an approximation of quantum theory. We are interested in the early phase of relativistic heavy-ion collisions experimentally studied at RHIC and the LHC. Within the Color Glass Condensate (CGC) approach, see {\it e.g.} the review articles \cite{Iancu:2003xm,Gelis:2012ri}, color charges of valence quarks of the colliding nuclei act as sources of long wavelength chromodynamic fields which can be treated as classical because of their large occupation numbers. The system of non-equilibrium gluon fields created in the nuclear collision is called {\it glasma}. At the earliest moment of the collision, the glasma is dominated by the chormoelectric and chromomagnetic fields parallel to the beam axis and later on transverse fields show up. 

It has been found numerically \cite{Romatschke:2005pm,Romatschke:2006nk}, see also \cite{Fukushima:2007yk}, that there is an unstable exponentially growing mode in the course of glasma's evolution. The mode was identified as the Weibel instability \cite{Wei59} which is well known in physics of electromagnetic plasma.  A presence of the chromodynamic Weibel instability in relativistic heavy-ion collisions was first argued in \cite{Mrowczynski:1993qm} and further on studied in detail, see the review \cite{Mrowczynski:2016etf}. 

The Weibel instability occurs when charged particles with anisotropic momentum distribution interact with the magnetic field generated by the particles. As explained in detail in  \cite{Mrowczynski:2016etf}, there is an energy transfer from the particles to the field which causes its exponential growth. In glasma there are no particles but high-frequency modes of classical fields are often treated as particles \cite{Romatschke:2005pm,Romatschke:2006nk}. 

The problem of unstable glasma was studied in the series of papers  \cite{Iwazaki:2007es,Iwazaki:2008xi,Fujii:2008dd,Fujii:2009kb} where a particular attention was paid to strong longitudinal chormoelectric and chromomagnetic fields generated at the earliest phase of nuclear collisions. It was suggested \cite{Iwazaki:2007es,Iwazaki:2008xi,Fujii:2008dd,Fujii:2009kb} that the unstable mode found in the numerical simulation \cite{Romatschke:2005pm,Romatschke:2006nk} is not the Weibel but rather Nielsen-Olesen instability \cite{Nielsen:1978rm} which occurs when spin 1 charged particles circulate in a uniform magnetic field. There was considered  \cite{Fujii:2008dd} a possible role of the vacuum instability due to strong electric field which according to the Schwinger mechanism \cite{Schwinger:1951nm} causes a spontaneous generation of particle-antiparticle pairs from vacuum. A stability of oscillatory chromomagnetic fields was also studied \cite{Berges:2011sb} in the context of glasma and a coexistence of the Nielsen-Olesen instability with the phenomenon of parametric resonance was found.

We intend to clarify what are the unstable modes of evolving glasma. Since the simulation \cite{Romatschke:2005pm,Romatschke:2006nk} was preformed in terms of classical fields we study a stability of classical field configurations. We start with the simplest case of  constant and uniform chromomagnetic and chromoelectric fields. The fields which are truly constant and uniform are obviously an idealization but our results are relevant for fields which are approximately constant and uniform in a limited space-time domain. 

Here we focus on a specific aspect of nonAbelian theory which has not been explored yet. The constant and homogeneous chromoelectric and chromomagnetic fields can occur due to the potentials which are of single color and as in electrodynamics linearly depend on coordinates. We call such configurations Abelian.  However, the fields can be also generated by the multi-color non-commuting potentials and then we have genuinely nonAbelian configurations. We note that the Abelian and nonAbelian configurations are physically inequivalent as they cannot be related to each other by a gauge transformation. It was also proved \cite{Brown:1979bv} that there are only these two gauge-inequivalent configurations which produce the space-time uniform chromodynamic fields. 

Stability of the Abelian configurations of constant and uniform chromoelectric and chromomagnetic fields was studied in \cite{Chang:1979tg,Sikivie:1979bq} and later on repeatedly analyzed, see {\it e.g.} \cite{Fujii:2008dd,Fujii:2009kb,Iwazaki:2008xi,Berges:2011sb}. However, the nonAbelian configurations seem to be more relevant for glasma. The point is that the chromoelectric and chromomagnetic fields $E_a, B_a$ from the earliest phase of the collisions are generated along the beam axis $z$ in a nonAbelian manner \cite{Chen:2015wia,Lappi:2006fp}. Specifically, the  $E_a, B_a$ fields occur due to the transverse pure gauge potentials of initial nuclei $A^i_{1a}, A^i_{2a}$ as $E_a = -g f^{abc} A^i_{1b} \, A^i_{2c}$ and $B_a = -g f^{abc}  \epsilon^{zij} A^i_{1b} \, A^j_{2c}$, where $f^{abc}$ are the structure constants of the ${\rm SU}(N_c)$ group and $\epsilon^{zij}$ is the totally asymmetric tensor. We are aware of only one paper \cite{Tudron:1980gq} where the stability of nonAbelian uniform configuration of chromomagnetic field was briefly discussed. A presence of an unstable mode was indicated but a complete spectrum of modes was not derived. 

We perform a comparative study of linear stability of Abelian and nonAbelian configurations of constant and homogeneous chromomagnetic and chromoelectric fields. We are mostly interested in the nonAbelian configurations but for a completeness of our study we repeat with minor refinements the stability analyses of Abelian configurations presented in \cite{Chang:1979tg,Sikivie:1979bq}. Throughout our whole study we use the background gauge while the axial gauges (different for the chromomagnetic and  chromoelectric configurations) were applied in \cite{Chang:1979tg,Sikivie:1979bq}. Using one gauge facilitates comparisons of various cases. 

We note that comparative analyses of the Abelian and nonAbelian configurations of uniform chromoelectric and chromomagnetic fields can be found in \cite{Brown:1979bv} and \cite{Friedman:1995iy}. A motion of a classical particle was shown to be significantly different in the two cases \cite{Brown:1979bv}. Quantum matter fields of spin 0 and 1/2  also behave differently in the background of Abelian and nonAbelian chromodynamic fields \cite{Brown:1979bv,Friedman:1995iy}. However, the self-interaction of nonAbelian fields, which is of our main interest, was not studied in \cite{Brown:1979bv} and \cite{Friedman:1995iy}.

Our paper is organized as follows. In Sec.~\ref{sec-linear-QCD} we present the linearized Yang-Mills equations in the background gauge which are subsequently used  in stability analyses. In Secs.~\ref{sec-Abelian-B} and \ref{sec-nonAbelian-B} we discuss, respectively, Abelian and nonAbelian configurations of the constant homogeneous chromomagnetic field. Secs.~\ref{sec-Abelian-E} and \ref{sec-nonAbelian-E} are devoted analogously to the constant homogeneous chromoelectric field. Our study is closed in Sec.~\ref{sec-final}. After summarizing our considerations, we briefly discuss the relevance of our results for fields which are uniform only in a limited spatial domain. Finally, we outline a perspective for further research. 

Throughout the paper the indices $i,j = x, y, z$ and $\mu, \nu = 0, 1, 2, 3$ label, respectively, the Cartesian spatial coordinates and those of Minkowski space. The signature of the metric tensor is $(+,-,-,-)$.  The indices $a, b = 1, 2, \dots N_c^2 -1$ numerate color components in the adjoint representation of SU($N_c$) gauge group. We neglect henceforth the prefix `chromo' when referring to chromoelectric or chromomagnetic fields. Since we study chromodynamics only, this should not be confusing.

\section{Linearized Classical Chromodynamics}
\label{sec-linear-QCD}

The Yang-Mills equations written in the adjoint representation of  the SU($N_c$) gauge group are
\be
\label{YM-eqs}
D^{ab}_\mu F_b^{\mu \nu} = j_a^\nu, 
\ee
where $D^{ab}_\mu \equiv \partial_\mu \delta^{ab} - g f^{abc}A^c_\mu$, $j_a^\nu$ is the color current and the strength tensor is 
\be
\label{F1}
F^{\mu\nu}_a=\partial^\mu A^\nu_a - \partial^\nu A^\mu_a + g f_{abc} A^\mu_b A^\nu_c .
\ee
The electric and magnetic fields are given as
\be
\label{EB-field}
E_a^i = F^{i0} , ~~~~~~~~~~
B_a^i = \frac{1}{2} \epsilon^{ijk} F_a^{kj} ,
\ee
where $\epsilon^{ijk}$ is the Levi-Civita fully antisymmetric tensor.

We assume that the potential $\bar{A}_a^\mu$ solves the Yang-Mills equation (\ref{YM-eqs}) and we consider small fluctuations $a_a^\mu$ around $\bar{A}_a^\mu$. So, we define the potential 
\be
\label{A=Abar+a}
A_a^\mu(t, {\bf r}) \equiv \bar{A}_a^\mu(t, {\bf r})  + a_a^\mu(t, {\bf r}),
\ee
such that $|\bar{A}(t, {\bf r})| \gg |a(t, {\bf r})|$.

Assuming that the background potential $\bar{A}^\mu_a$ satisfies the Lorentz gauge condition $\partial_\mu \bar{A}^\mu_a = 0$ while the fluctuation potential $a_a^\mu$ that of the background gauge 
\be
\label{background-gauge}
\bar{D}^{ab}_\mu a_b^\mu = 0,
\ee
where $\bar{D}^{ab}_\mu \equiv \partial_\mu \delta^{ab} - g f^{abc}\bar{A}^c_\mu$, the Yang-Mills equation linearized in $a_a^\mu$ can be written as
\be
\label{linear-YM-background}
\big[ g^{\mu \nu}(\bar{D}_\rho \bar{D}^\rho)_{ac} + 2 g f^{abc} \bar{F}_b^{\mu\nu} \big] a^c_\nu = 0.
\ee

The background gauge appears particularly convenient for our purposes because different color and space-time components of $a^a_\mu$ are mixed only through the tensor $\bar{F}_b^{\mu\nu}$ which enters Eq.~(\ref{linear-YM-background}). In case of other gauges, {\it e.g.} the Lorentz gauge $\partial_\mu a^\mu_a = 0$, the mixing is more severe. 

Throughout our analysis, which includes Abelian and nonAbelian configurations of magnetic and electric fields, we use the background gauge which facilitates comparisons of various cases. However, our further considerations are limited to the SU(2) gauge group when $ f^{abc} = \epsilon^{abc}$ with $a,b = 1, 2, 3$.

\section{Abelian Configuration of Magnetic Field}
\label{sec-Abelian-B}

The constant homogeneous magnetic field along the axis $x$ occurs for a potential $\bar{A}^i_a$ which is known from the Abelian theory. Specifically, $\bar{\bf A}_a(t, {\bf r}) = \delta^{a1}(0,0,yB)$, where $B$ is a constant and ${\bf r}=(x,y,z)$. Then,  using Eqs.~(\ref{EB-field}), one finds  ${\bf E}_a(t, {\bf r}) =0$ and ${\bf B}_a(t, {\bf r}) = \delta^{a1}(B,0,0)$. We also note that the only non-vanishing components of the strength tensor are $\bar{F}_1^{zy} = - \bar{F}_1^{yz} = B$. The Abelian configuration solves the Yang-Mills equations (\ref{YM-eqs}) with vanishing current $j^\mu_a$. The nonAbelian terms disappear because there is only one color component. We also note that the chosen potential satisfies the Lorentz gauge condition.

When $\bar{A}^\mu_a(t, {\bf r}) = \delta^{a1}(0,0,0,yB)$, the equation (\ref{linear-YM-background}) of $a^\mu_a$ becomes
\be
\label{a-mu-bcg}
\Box a^\mu_a - 2 g By \epsilon^{ab1} \partial_z a^\mu_b 
- 2g B \epsilon^{ab1} ( \delta^{\mu y} a^z_b - \delta^{\mu z} a^y_b )
- g^2 B^2 y^2 \epsilon^{ac1} \epsilon^{cb1} a^\mu_b = 0 .
\ee
One sees that the color component $a^\mu_1$ decouples from the remaining two color components and it satisfies the free equation of motion. So, the functions $a^\mu_1$ represent free waves which we do not consider anymore.

Defining the functions 
\ba
\begin{split}
& T^\pm = a^0_2 \pm i a^0_3 , ~~~~~
    X^\pm = a^x_2 \pm i a^x_3 , 
\\[2mm]
& Y^\pm = a^y_2 \pm i a^y_3 , ~~~~~
   Z^\pm = a^z_2 \pm i a^z_3 ,
\end{split}
\ea
the equation (\ref{a-mu-bcg}) provides 
\ba
\label{T-bcg}
&&\big(\Box  \pm 2i g By \partial_z  
+ g^2 B^2 y^2 \big) T^\pm = 0 ,
\\[2mm]
\label{X-bcg}
&&\big(\Box \pm 2i g By \partial_z  
+ g^2 B^2 y^2 \big) X^\pm = 0 ,
\\[2mm]
\label{Y-bcg}
&&\big(\Box \pm 2i g By \partial_z 
+ g^2 B^2 y^2 \big) Y^\pm 
\pm 2i g B Z^\pm = 0 ,
\\[2mm]
\label{Z-bcg}
&&\big(\Box \pm 2i g By \partial_z  
+ g^2 B^2 y^2 \big) Z^\pm 
\mp 2i g B Y^\pm = 0 .
\ea
The equations of $T^\pm$ and $X^\pm$ have the diagonal form. To diagonalize the equations of $Y^\pm$ and $Z^\pm$ one defines the functions 
\be
U^\pm \equiv Y^+ \pm iZ^+ ,
~~~~~~~~~~~~
W^\pm \equiv Y^- \pm iZ^- ,
\ee
which allow one to change the equations (\ref{Y-bcg}) - (\ref{Z-bcg}) into 
\ba
\label{U-bcg}
&&\big(\Box + 2i g By \partial_z 
\pm 2g B + g^2 B^2 y^2 \big) U^\pm = 0 ,
\\[2mm]
\label{W-bcg}
&&\big(\Box - 2i g By \partial_z  
\mp 2 g B + g^2 B^2 y^2 \big) W^\pm = 0  .
\ea

We assume that the functions $a^\mu_a$ depend on $t, x, z$ as $e^{-i(\omega t - k_x x - k_z z)}$. Since the functions should be real, only their real parts are of physical meaning. Now, the equations (\ref{T-bcg}), (\ref{X-bcg}) and (\ref{U-bcg}), (\ref{W-bcg}) read
\ba
\label{Eq-T}
\Big(-\omega^2 + k_x^2 + (k_z \mp g B y)^2  - \frac{d^2}{dy^2} \Big) T^\pm(y) &=& 0 ,
\\[2mm] 
\label{Eq-X}
\Big(-\omega^2 + k_x^2 
+ ( k_z \mp  g yB )^2 - \frac{d^2}{dy^2} \Big) X^\pm (y) &=& 0 ,
\\[2mm] 
\label{Eq-U} 
\Big(-\omega^2  \pm 2gB + k_x^2 
+ ( k_z -  g yB )^2 - \frac{d^2}{dy^2}\Big) U^\pm(y) &=& 0 ,
\\[2mm] 
\label{Eq-W} 
\Big(-\omega^2  \mp 2gB + k_x^2 
+ ( k_z +  g yB )^2 - \frac{d^2}{dy^2}\Big) W^\pm(y) &=& 0 .
\ea
We note that one obtains exactly the same equations (\ref{Eq-X}), (\ref{Eq-U}) and (\ref{Eq-W}) using the temporal axial gauge $a_a^0=0$ which was applied in Refs.~\cite{Chang:1979tg,Sikivie:1979bq}.

Since the eigenenergy Schr\"odinger equation of harmonic oscillator can be written as
\be
\label{Schr-eq-HO}
\Big(- 2m{\cal E} + m^2 \bar\omega^2(y - y_0)^2 - \frac{d^2}{dy^2} \Big) \varphi(y) = 0 ,
\ee
where $m$ is the oscillator mass, ${\cal E}$ its energy and $\bar\omega$ is the frequency of the corresponding classical oscillator, one observes that Eqs.~(\ref{Eq-T}) - (\ref{Eq-W}) coincide with Eq.~(\ref{Schr-eq-HO}) under the following replacements 
\be
\omega^2 + d - k_x^2 ~ \rightarrow ~2m{\cal E},  ~~~~~~~~~~ 
g B ~ \rightarrow ~ m \bar\omega , ~~~~~~~~~~
\pm \frac{k_z}{gB} ~ \rightarrow ~ y_0 ,
\ee
where $d = 0$ for Eqs.~(\ref{Eq-T}), (\ref{Eq-X}) and $d = \mp 2gB$ for Eqs.~(\ref{Eq-U}), (\ref{Eq-W}). 

Since ${\cal E} = \bar\omega  (n + 1/2)$ with $n = 0,\,1,\,2,\, \dots $, the frequency squared $\omega^2$ is 
\be
\label{omega2-1}
\omega_0^2 = 2gB \Big(n + \frac{1}{2}\Big) + k_x^2 ,  ~~~~~~ n = 0,\,1,\,2,\, \dots 
\ee 
for Eqs.~(\ref{Eq-T}), (\ref{Eq-X}) and
\be
\label{omega2-2}
\omega_\pm^2 = 2gB \Big(n + \frac{1}{2}\Big) \pm 2gB + k_x^2,  ~~~~~~ n = 0,\,1,\,2,\, \dots 
\ee
for Eqs.~(\ref{Eq-U}), (\ref{Eq-W}). It should be stressed that although we refer to the Schr\"odinger equation to find the frequencies (\ref{omega2-1}), (\ref{omega2-2}) the solutions are purely classical - the Planck constant $\hbar$ does not show up in the final formulas. The frequencies squared (\ref{omega2-1}), (\ref{omega2-2}) are `quantized' because the fluctuation field $a^\mu_a$ is assumed to be limited everywhere. This is analogous to the requirement that a wave function, which solves the  Schr\"odinger equation, is normalizable. 

One sees that $\omega_0^2 \ge 0$ and $\omega_+^2 \ge 0$ for any $n$ but $\omega_-^2 = - gB + k_x^2$ for $n=0$ and consequently, it is negative for $k_x^2 < gB$. Then, there are unstable modes of $U^-$ and $W^+$ which grow as $e^{\gamma t}$ with $\gamma \equiv \sqrt{gB - k_x^2}$. This is the well-known Nielsen-Olesen instability \cite{Nielsen:1978rm}. The unstable modes are paired with the overdamped modes which decrease in time as $e^{-\gamma t}$.

Let us now discuss a character of the solutions of Eqs.~(\ref{Eq-T}) - (\ref{Eq-W}). The potentials $a^\mu_a$ are assumed to depend on time as $e^{i\omega  t}$ but to see how a given combination of $a^\mu_a$ evolves in time, one should consider only the real parts of $a^\mu_a$. 

\vspace{3mm}
\underline{Modes $T^\pm$ and  $X^\pm$}
\vspace{1mm}

The modes $T^\pm$ and $X^\pm$ are stable. Assuming that  $T^+ \not= 0$ while $T^-= X^\pm=U^\pm = W^\pm = 0$, one finds that $T^+$ represents the wave which rotates in the two-dimensional color space spanned by the colors 2 and 3. The mode $T^-$ is similar but it rotates in the opposite direction than $T^+$. The modes $X^\pm$ behave as $T^\pm$. 

\vspace{3mm}
\underline{Modes $U^\pm$ and $W^\pm$}
\vspace{1mm}

The modes $U^+$ and $W^-$ are always stable. When $U^+ \not=0$ and $U^- = X^\pm = W^\pm = 0$, one finds that $U^+$ represents the wave which rotates in both two-dimensional 2-3 color and $y$-$z$ coordinate spaces. There is analogous situation with the stable modes $W^-$. The modes $U^-$ and $W^+$ can be stable or unstable. If the modes are stable, they are similar to $U^+$ and $W^-$. In case of unstable and overdamped modes $U^-$ and $W^+$, which depend on time as $e^{\gamma t}$ and $e^{- \gamma t}$, the small field wave does not rotate neither in color nor in coordinate space. 

\section{NonAbelian Configuration of Magnetic Field}
\label{sec-nonAbelian-B}

A nonAbelian configuration of $\bar{A}^i_a$ which produces a constant homogeneous magnetic field ${\bf B}_a = \delta^{a1}(B,0,0)$ can be chosen as 
\ba
\label{Abar-matrix}
\bar{A}^\mu_a = 
\left[ {\begin{array}{cccc}
0 ~~~~& 0 & 0 & 0 \\
0 ~~~~& 0 & 0 & \sqrt{B/g} \\
0 ~~~~& 0 & \sqrt{B/g} & 0 \\
 \end{array} } \right]  ,
\ea
where the Lorentz index $\mu$ numerates the columns and the color index $a$ numerates the rows. The potential (\ref{Abar-matrix}), which obviously satisfies the Lorentz gauge condition, does not solve the Yang-Mills equation (\ref{YM-eqs}) with $j_a^\mu=0$. Instead one gets
\ba
\label{eq-j-B-matrix}
\left[ {\begin{array}{cccc}
0 ~~~~& 0 & 0 & 0 \\
0 ~~~~& 0 & 0 & g^{1/2} B^{3/2} \\
0 ~~~~& 0 & g^{1/2} B^{3/2} & 0 \\
 \end{array} } \right] = j^\mu_a .
\ea
Following \cite{Tudron:1980gq}, we assume that the current, which enters the Yang-Mills equation, equals the left-hand side of Eq.~(\ref{eq-j-B-matrix}). Then, the potential (\ref{Abar-matrix}) solves the Yang-Mills equations (\ref{YM-eqs}). 

The equation of motion of the small field $a^\mu_a$ (\ref{linear-YM-background}) is found to be
\be
\label{a-mu-bcg-nA}
\Box a^\mu_a 
+ 2g A (\epsilon^{a3b}\partial_y + \epsilon^{a2b}\partial_z ) a^\mu_b
- g^2 A^2 (\epsilon^{a2e}\epsilon^{e2b} + \epsilon^{a3e}\epsilon^{e3b} ) a^\mu_b
+ 2g^2 A^2 \epsilon^{a1b}( \delta^{\mu y} a^z_b -  \delta^{\mu z} a^y_b) 
 = 0 ,
\ee
where $A \equiv \sqrt{B/g}$.

Assuming that $a_a^\mu (t,x,y,z) = e^{-i (\omega t - {\bf k}\cdot {\bf r})} a_a^\mu$, where ${\bf k} = (k_x, k_y, k_z)$ and  ${\bf r} = (x,y,z)$, Eqs.~(\ref{a-mu-bcg-nA}) are changed into the following set of algebraic equations
\ba
\label{eq-Bt}
\hat{M}_B^t \,\vec{a^t} &=& 0,
\\
\label{eq-Bx}
\hat{M}_B^x \,\vec{a^x} &=& 0,
\\
\label{eq-Byz}
\hat{M}_B^{yz} \,\vec{a^{yz}} &=& 0,
\ea
where 
\ba
\label{matrix-MBtx}
\hat{M}_B^t  = \hat{M}_B^x = 
\left[ {\begin{array}{ccc}
-\omega^2 + {\bf k}^2 + 2 g^2 A^2 & - 2ig A k_y & 2ig A k_z   \\[2mm]
2ig A k_y & -\omega^2 + {\bf k}^2  + g^2 A^2 & 0  \\[2mm]
-2ig A k_z & 0 & -\omega^2 + {\bf k}^2 +  g^2 A^2 \\
 \end{array} } \right]  ,
\ea
\ba
\nn
\hat{M}_B^{yz}  =
\left[ {\begin{array}{cccccc}
-\omega^2 + {\bf k}^2 + 2 g^2 A^2 & - 2ig A k_y & 2ig A k_z  & 0 & 0 & 0 \\[2mm]
2ig A k_y & -\omega^2 + {\bf k}^2  + g^2 A^2 & 0 & 0 & 0 & -  2 g^2 A^2 \\[2mm]
-2ig A k_z & 0 & -\omega^2 + {\bf k}^2 +  g^2 A^2 & 0 & 2 g^2 A^2 & 0 \\[2mm]
0 & 0 & 0 & -\omega^2 + {\bf k}^2 + 2 g^2 A^2  & - 2ig A k_y &  2ig A k_z  \\[2mm]
0 & 0 & 2 g^2 A^2 & 2ig A k_y & -\omega^2 + {\bf k}^2 + g^2 A^2 & 0 \\[2mm]
0 & -  2 g^2 A^2 & 0 & -2ig A k_z & 0 &  -\omega^2 + {\bf k}^2 + g^2 A^2 \\
 \end{array} } \right] , 
\\ \label{matrix-MByz}
\ea
and 
\ba
\vec{a^t}  =  \left[ {\begin{array}{c}
a^t_1 \\[2mm]
a^t_2 \\[2mm]
a^t_3 \\
 \end{array} } \right],
~~~~~~~~~~
\vec{a^x}  =  \left[ {\begin{array}{c}
a^x_1 \\[2mm]
a^x_2 \\[2mm]
a^x_3 \\
 \end{array} } \right],
~~~~~~~~~~
\vec{a^{yz}}  =  \left[ {\begin{array}{c}
a^y_1 \\[2mm]
a^y_2 \\[2mm]
a^y_3 \\[2mm]
a^z_1 \\[2mm]
a^z_2 \\[2mm]
a^z_3 \\
 \end{array} } \right] .
\ea

Since the equations (\ref{eq-Bt}),  (\ref{eq-Bx}) and  (\ref{eq-Byz}) are all homogeneous, they have solutions if
 \be
{\rm det}\hat{M}_B^t=0, ~~~~~~~~~~
{\rm det}\hat{M}_B^x=0, ~~~~~~~~~~
{\rm det}\hat{M}_B^{yz}=0 ,
\ee
which are the dispersion equations. 

\begin{figure}[t]
\begin{minipage}{87mm}
\centering
\includegraphics[scale=0.23]{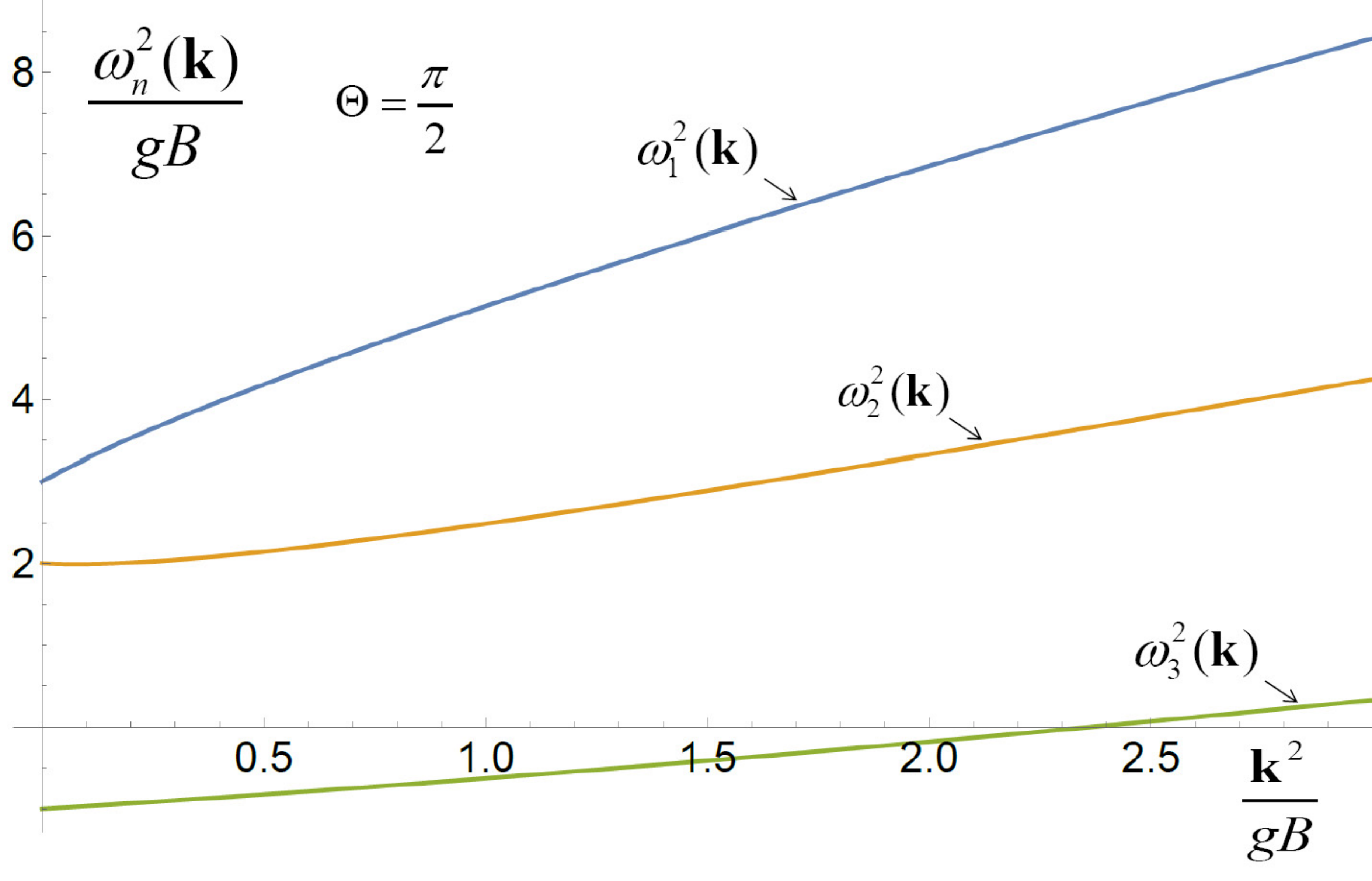}
\vspace{-6mm}
\caption{$\omega_n^2({\bf k})$ as a function of ${\bf k}^2$ for $\Theta = \pi/2$.}
\label{FigB-yz-1}
\end{minipage}
\hspace{3mm}
\begin{minipage}{87mm}
\centering
\vspace{-1mm}
\includegraphics[scale=0.23]{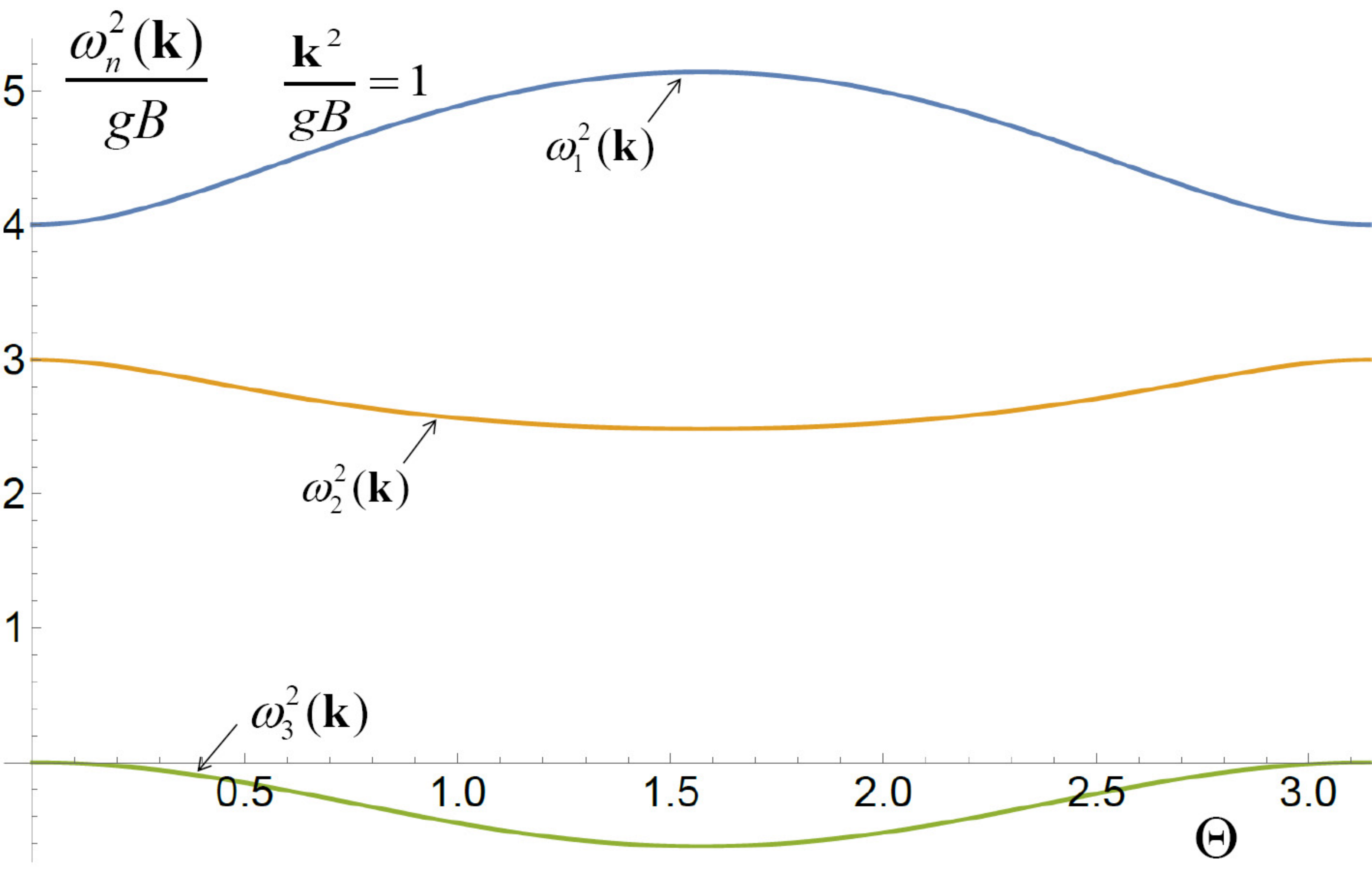}
\vspace{-3mm}
\caption{$\omega_n^2({\bf k})$ as a function of $\Theta$ for ${\bf k}^2=gB$.}
\label{FigB-yz-2}
\end{minipage}
\end{figure}

\subsection{Equations ${\rm det}\hat{M}_B^t=0$ and ${\rm det}\hat{M}_B^x=0$}

Computing the determinant of the matrix $\hat{M}_B^t$, the dispersion equation ${\rm det}\hat{M}_B^t=0$ becomes
\ba
 (-\omega^2 + k^2 +  g B) 
 \Big(\omega^4 -\omega^2(2 k^2 + 3gB)  
+ k^4 + gB (3 k^2 - 4k_T^2) 
+ 2g^2 B^2 \Big) = 0,
\ea
where $k \equiv |{\bf k}|$ and $k_T \equiv \sqrt{k_y^2 + k_z^2}$. The solutions are
\be
\label{sol-nonAbelian-x-TAG}
\omega_\pm^2({\bf k}) = \frac{1}{2} \Big( 2 k^2 + 3 gB
\pm \sqrt{g^2 B^2+ 16 gB k_T^2 }~\Big),
~~~~~~~~~~~~~~
\omega_0^2({\bf k}) = g B + k^2  .
\ee
One observes that $\omega_\pm^2({\bf k})$ and $\omega_0^2({\bf k})$ are always positive. Consequently the modes $\pm \omega_+({\bf k})$, $\pm \omega_-({\bf k})$ and $\pm \omega_0({\bf k})$ are real and stable. The solutions of the equation ${\rm det}\hat{M}_B^x=0$ are obviously the same as those of ${\rm det}\hat{M}_B^t=0$.

The waves represented by the solutions of the equations ${\rm det}\hat{M}_B^x=0$ and ${\rm det}\hat{M}_B^t=0$ rotate not in the two-dimensional color subspace, as the analogous solutions of the Abelian configuration, but in the three-dimensional space. 

\subsection{Equation ${\rm det}\hat{M}_B^{yz}=0$}

Computing the determinant of the matrix $\hat{M}_B^{yz}$ as
\ba
\label{det-MByz}
{\rm det}\hat{M}_B^{yz} =
\big[-6 g^6 A^6 + (k^2 - \omega^2)^3 
+ g^4 A^4  (k^2 - 4 k_T^2 - \omega^2) 
+ 4 g^2 A^2 (k^2 - \omega^2) (k^2 - k_T^2 - \omega^2)\big]^2,
\ea 
one finds that the dispersion equation ${\rm det}\hat{M}_B^{yz}=0$ is cubic in $x\equiv \omega^2$ and it reads
\be
\label{cubic-eq}
x^3 + a_2x^2 + a_1 x + a_0 = 0,
\ee
with 
\ba
\label{a2-B}
a_2 &\equiv& - 4 g^2 A^2 - 3 k^2 ,
\\
\label{a1-B}
a_1 &\equiv& g^4 A^4  + 8 g^2 A^2 k^2 - 4 g^2 A^2 k_T^2 + 3 k^4 ,
\\
\label{a0-B}
a_0 &\equiv& 6 g^6 A^6 - g^4 A^4  k^2 + 4 g^4 A^4 k_T^2 - 4 g^2 A^2 k^2(k^2 - k_T^2) - k^6 .
\ea
We note that because of the square in the determinant (\ref{det-MByz}) each solution of the cubic equation is doubled. 

As well known, see {\it e.g.} \cite{Bronshtein-Semendyayev-1985}, all three roots of a cubic equation can be found algebraically. Since the coefficients  $a_0, \, a_1, \, a_2$ are real, the character of the roots depends on a value of the discriminant 
\ba
\label{discriminant-def}
\Delta=18 \, a_0 a_1 a_2  - 4 \, a_2^3 a_0 + a_1^2 a_2^2 - 4 \, a_1^3 - 27 \, a_0^2 .
\ea
One distinguishes three cases:
\begin{itemize}

\item if $\Delta > 0$, the roots are real and distinct;

\item if $\Delta = 0$, the roots are real and at least two of them coincide;

\item if $\Delta < 0$,  one root is real and the remaining two are complex.

\end{itemize}

The discriminant (\ref{discriminant-def}) with the coefficients (\ref{a2-B}), (\ref{a1-B}), (\ref{a0-B}) is computed as
\ba
\label{discriminant-B}
\frac{\Delta}{16g^3B^3} &=& 9 g^3 B^3 + 68 g^2 B^2 k_T^2 + 49 g B k_T^4  +  16 k_T^6.
\ea 
As seen, $\Delta > 0$ and there are three distinct real solutions of the equation ${\rm det}\hat{M}_B^{yz}=0$. 

The real solutions of the cubic equation (\ref{cubic-eq}) can be written down in the Vi\` ete's trigonometric form \cite{Bronshtein-Semendyayev-1985}
\be
\label{x-n}
x_n = 
2 \sqrt{\frac{-p}{3}} \cos\bigg[\frac{1}{3}\arccos\Big(\frac{3 q}{2p} \sqrt{\frac{-3}{p}}~\Big) - \frac{2\pi (n-1)}{3}\bigg]
- \frac{a_2}{3} ,
\ee
where $n=1,2,3$ and 
\be
\label{p-q}
p \equiv \frac{3 a_1 - a_2^2}{3},
~~~~~~~~~~~~~~
q \equiv \frac{2 a_2^3 - 9 a_2 a_1 + 27 a_0}{27}  .
\ee
These formulas assume that $p<0$ and that the argument of the arccosine belongs to $[-1,1]$. These conditions are guaranteed as long as $\Delta > 0$ which is the case under consideration.

\begin{figure}[t]
\begin{minipage}{87mm}
\centering
\vspace{-2mm}
\includegraphics[scale=0.37]{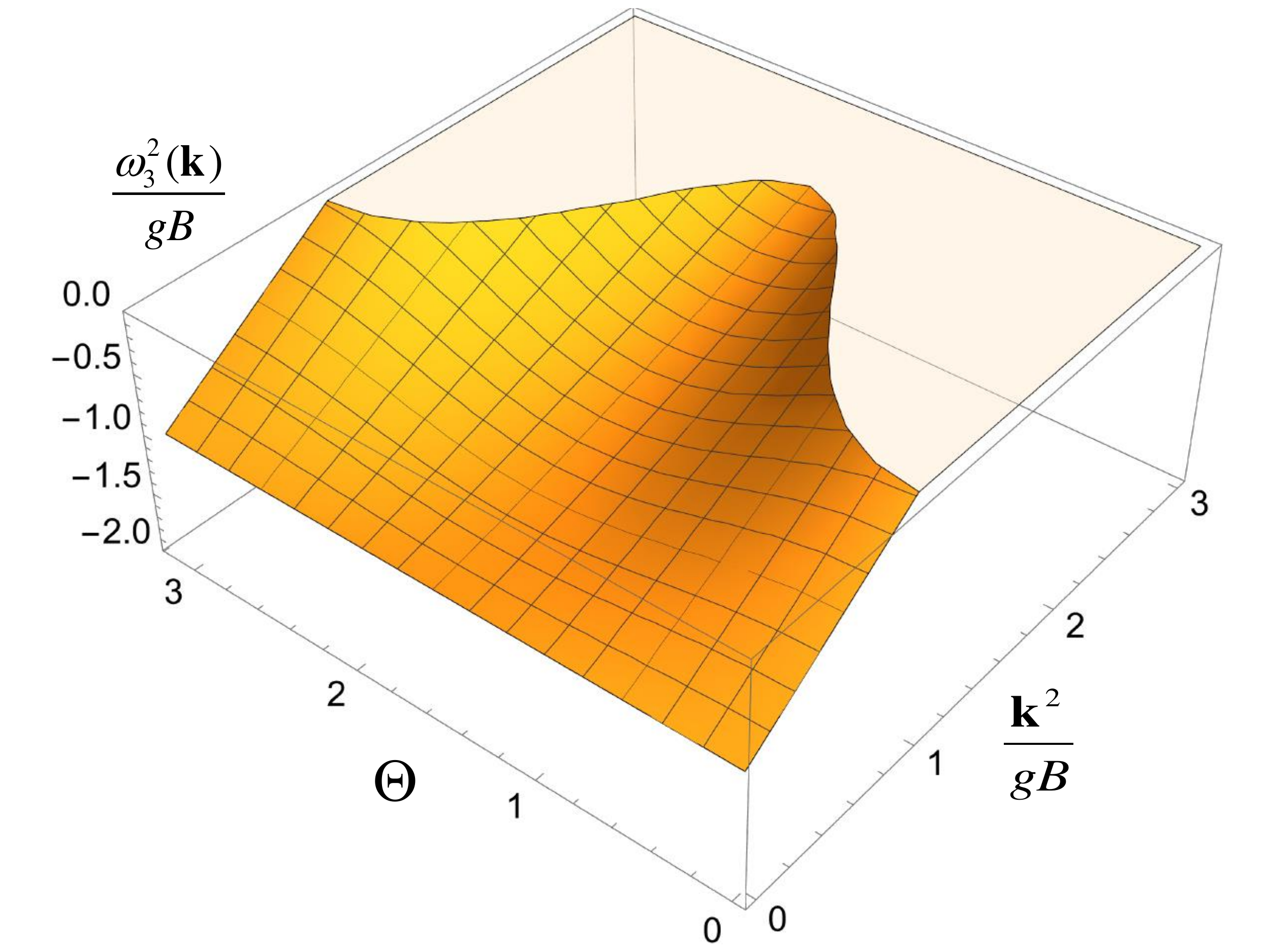}
\end{minipage}
\hspace{2mm}
\begin{minipage}{87mm}
\centering
\vspace{0mm}
\includegraphics[scale=0.37]{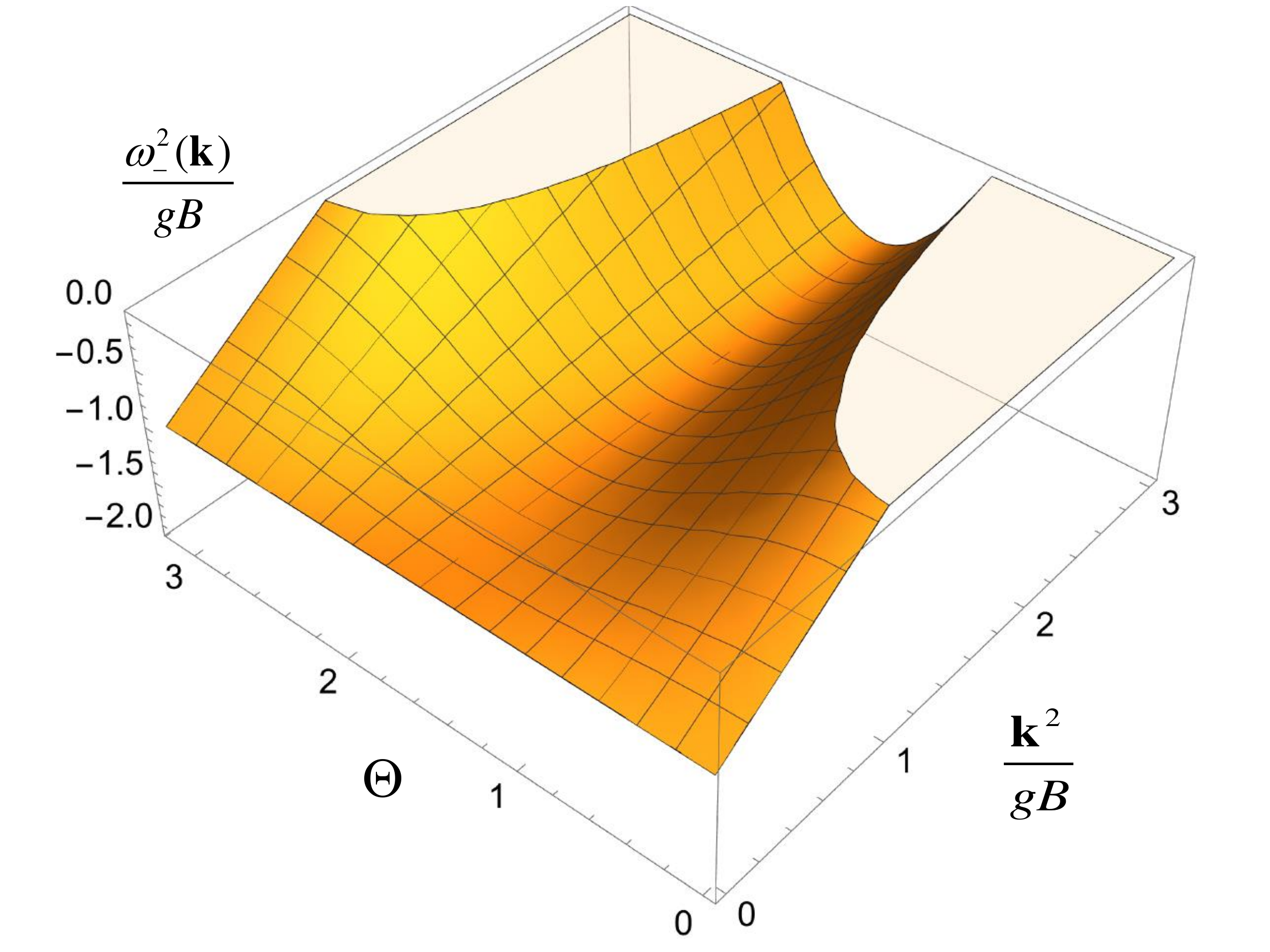}
\end{minipage}
\vspace{-3mm}
\caption{Unstable modes in the nonAbelian (left panel) and Abelian (right panel) configurations as a function of ${\bf k}^2$ and $\Theta$.}
\label{FigB-unstable}
\end{figure}

We show $\omega_n^2({\bf k})$ with $n=1,2,3$ as a function of $k^2$ for $\Theta = \pi/2$ in Fig.~\ref{FigB-yz-1}  and $\omega_n^2({\bf k})$ as a function of $\Theta$ for $k^2=gB$ in Fig.~\ref{FigB-yz-2}. The angle $\Theta$ defines the orientation of the wave vector ${\bf k}$ with respect to the magnetic field along the axis $x$. Therefore, $k_T = k \sin\Theta$. One observes that  $\omega_1^2({\bf k})$ and $\omega_2^2({\bf k})$ are everywhere positive and the corresponding modes $\pm \omega_1({\bf k})$ and $\pm \omega_2({\bf k})$ are stable. There is a domain shown in the left panel of Fig.~\ref{FigB-unstable} where $\omega_3^2({\bf k})$ is negative. For comparison we show in the right panel of Fig.~\ref{FigB-unstable} the domain of instability of the Abelian configuration discussed in Sec.~\ref{sec-Abelian-B}. The Abelian mode depends only on $k_x = k \cos\Theta$. One observes that the domain of instability of the Abelian configuration extends to infinity for a wave vector which is perpendicular to the magnetic field. In case of nonAbelian configuration, the domain of instability is limited for any orientation of the wave vector.  We note that with the pure imaginary unstable modes which exponentially grow in time there are paired overdamped modes which exponentially decay in time. 

The waves represented by the solutions of the equation ${\rm det}\hat{M}_B^{yz}=0$ rotate in the $y$-$z$ plane, as the analogous solutions of the Abelian configuration, but the rotation in the color space is not in the two-dimensional subspace but in the three-dimensional space. 

\section{Abelian Configuration of Electric Field}
\label{sec-Abelian-E}

The constant homogeneous electric field along the axis $x$ occurs for a potential $\bar{A}^i_a$ which is known from the Abelian theory. Specifically, $\bar{A}_a^\mu(t, {\bf r}) = \delta^{a1}(-xE,0,0,0)$, where $E$ is a constant and ${\bf r}=(x,y,z)$. Then, using Eqs.~(\ref{EB-field}), one finds  ${\bf E}_a(t, {\bf r}) = \delta^{a1}(E,0,0)$ and ${\bf B}_a(t, {\bf r}) =0$. We also note that the only non-vanishing elements of $\bar{F}_a^{\mu \nu}$ are $\bar{F}_1^{x0} = - \bar{F}_1^{0x} = E$. The chosen potential solves the Yang-Mills equations (\ref{YM-eqs}) with vanishing current. The nonAbelian terms disappear because there is only one color component. We also note that the chosen potential satisfies the Lorentz gauge condition.

When $\bar{A}^\mu_a(t, {\bf r}) = \delta^{a1}(-xE,0,0,0)$, the equation (\ref{linear-YM-background}) of $a^\mu_a$ becomes
\be
\label{a-mu-E-a}
\Box a^\mu_a - 2gEx \epsilon^{a1b} \partial_0 a^\mu_b 
+ 2gE \epsilon^{a1b} (\delta^{\mu 0} a^x_b + \delta^{\mu x} a^0_b)  
+ g^2 E^2 x^2  \epsilon^{a1d} \epsilon^{d1b} a^\mu_b = 0 .
\ee
One sees that the color component $a^\mu_1$ decouples from the remaining two color components and it satisfies the free equation of motion. So, the functions $a^\mu_1$ represent free waves which we do not consider any more.

Defining the functions
\ba
\begin{split}
& T^\pm(x) = a^0_2(x) \pm i a^0_3(x) , ~~~~~
     X^\pm(x) = a^x_2(x) \pm i a^x_3(x) , 
\\[2mm]
& Y^\pm(x) = a^y_2(x) \pm i a^y_3(x) , ~~~~~
     Z^\pm(x) = a^z_2(x) \pm i a^z_3(x) , 
\end{split}
\ea
Eqs.~(\ref{a-mu-E-a}) provides the equations
\ba
\label{T-bgk}
&&\big(\Box  \mp 2i gEx \partial_0   - g^2 E^2 x^2 \big) T^\pm 
\pm 2igE X^\pm = 0 ,
\\[2mm]
\label{X-bgk}
&&\big(\Box \mp 2igEx \partial_0 - g^2 E^2 x^2 \big) X^\pm
\pm 2igE T^\pm  = 0 ,
\\[2mm]
\label{Y-bgk}
&&\big(\Box \mp 2igEx \partial_0  - g^2 E^2 x^2 \big) Y^\pm = 0,
\\[2mm]
\label{Zplus-bgk}
&&\big(\Box \mp 2igEx \partial_0  - g^2 E^2 x^2 \big) Z^\pm = 0.
\ea
The equations of $Y^\pm$ and $Z^\pm$ have a diagonal form while the equations of $T^\pm$ and $X^\pm$ are diagonalized using 
\be
G^\pm \equiv T^+ \pm X^+,
~~~~~~~~~~~~~~~~~~
H^\pm \equiv T^- \pm X^-.
\ee
Then, Eqs. (\ref{T-bgk}) and (\ref{X-bgk}) provide
\ba
\label{G-bgk}
&&\big(\Box  - 2i gEx \partial_0  \pm 2igE - g^2 E^2 x^2 \big) G^\pm = 0 ,
\\[2mm]
\label{H-bgk}
&&\big(\Box  + 2igEx \partial_0 \mp 2igE - g^2 E^2 x^2 \big) H^\pm = 0 .
\ea

Assuming that the functions $G^\pm, H^\pm, Y^\pm, Z^\pm$ depend on $t, y, z$ as  $e^{-i(\omega t - k_y y - k_z z)}$ we find
\ba 
\label{G-pm-bgk-k2}
&& \Big( k_y^2 + k_z^2 \pm 2i g E   - (\omega + g E x)^2 
- \frac{d^2}{dx^2} \Big) G^\pm(x) = 0,
\\[2mm]
\label{H-pm-bgk-k2}
&& \Big(k_y^2 + k_z^2 \mp 2i g E - (\omega - g E x)^2 
- \frac{d^2}{dx^2} \Big) H^\pm(x) = 0,
\\[2mm]
\label{Y-pm-bgk-k2}
&& \Big(k_y^2 + k_z^2 - (\omega \pm g E x)^2 - \frac{d^2}{dx^2} \Big) Y^\pm(x) = 0,
\\[2mm]
\label{Z-pm-bgk-k2}
&& \Big(k_y^2 + k_z^2 - (\omega \mp g E x)^2 - \frac{d^2}{dx^2} \Big) Z^\pm(x) = 0 .
\ea
We note that one obtains exactly the same equations (\ref{G-pm-bgk-k2}), (\ref{H-pm-bgk-k2}) and (\ref{Y-pm-bgk-k2}) using the axial gauge $a_a^z=0$ which was applied in Ref.~\cite{Chang:1979tg}.

Since the eigenenergy Schr\"odinger equation of inverted harmonic oscillator can be written as
\be
\label{Schr-eq-IHO}
\Big(- 2m{\cal E} - m^2 \bar\omega^2(x -  x_0)^2 - \frac{d^2}{dx^2} \Big) \varphi(x) = 0 ,
\ee
one sees that Eqs.~(\ref{G-pm-bgk-k2}) - (\ref{Z-pm-bgk-k2}) coincide with Eq.~(\ref{Schr-eq-IHO}) under the following replacements 
\be
k_y^2 + k_z^2 + d ~ \rightarrow ~ - 2m{\cal E}, ~~~~~~~~~~
g E ~ \rightarrow ~ m \bar\omega ,  ~~~~~~~~~~ 
\pm \frac{\omega}{gE} ~ \rightarrow ~  y_0, 
\ee
where $d=0$ for Eqs.~(\ref{Y-pm-bgk-k2}) and (\ref{Z-pm-bgk-k2}) and $d= \pm 2igE$ for Eqs.~(\ref{G-pm-bgk-k2}) and (\ref{H-pm-bgk-k2}). In the latter case we deal with the Schr\"odinger equation of non-Hermitian Hamiltonian. 

As discussed in detail in \cite{Barton:1984ey}, there are no normalizable solutions of the Schr\"odinger equation of inverted harmonic oscillator which reflects the fact that the solutions run away either to plus or minus infinite. In this sense the configuration of constant electric field is genuinely unstable. 

\section{NonAbelian Configuration of Electric Field}
\label{sec-nonAbelian-E}

A nonAbelian configuration of $\bar{A}^i_a$, which produces a constant homogeneous electric field ${\bf E}_a = \delta^{a1}(E,0,0)$, can be chosen as 
\ba
\label{pot-nonabel}
\bar{A}_a^\mu = 
\left[ {\begin{array}{cccc}
0 & 0 & 0 & 0 \\
\sqrt{E/g} & 0 & 0 & 0 \\
0 & \sqrt{E/g} & 0 & 0 \\
  \end{array} } \right] ,
\ea
where the Lorentz index $\mu$ numerates the columns and the color index $a$ numerates the rows. The potential (\ref{pot-nonabel}), which obviously satisfies the Lorentz gauge condition, does not solve the Yang-Mills equation (\ref{YM-eqs}) with $j_a^\mu=0$. Instead one gets
\ba
\label{eq-j-E-matrix}
\left[ {\begin{array}{cccc}
0 & 0 & 0 & 0 \\
g^{1/2} E^{3/2} & 0 & 0 & 0 \\
0 & - g^{1/2} E^{3/2} & 0 & 0 \\
 \end{array} } \right] = j^\mu_a .
\ea
Following \cite{Tudron:1980gq}, we assume that the current, which enters the Yang-Mills equation, equals the left-hand side of Eq.~(\ref{eq-j-E-matrix}).  Then, the potential (\ref{pot-nonabel}) solves the Yang-Mills equations (\ref{YM-eqs}). 

\begin{figure}[t]
\begin{minipage}{87mm}
\centering
\includegraphics[scale=0.26]{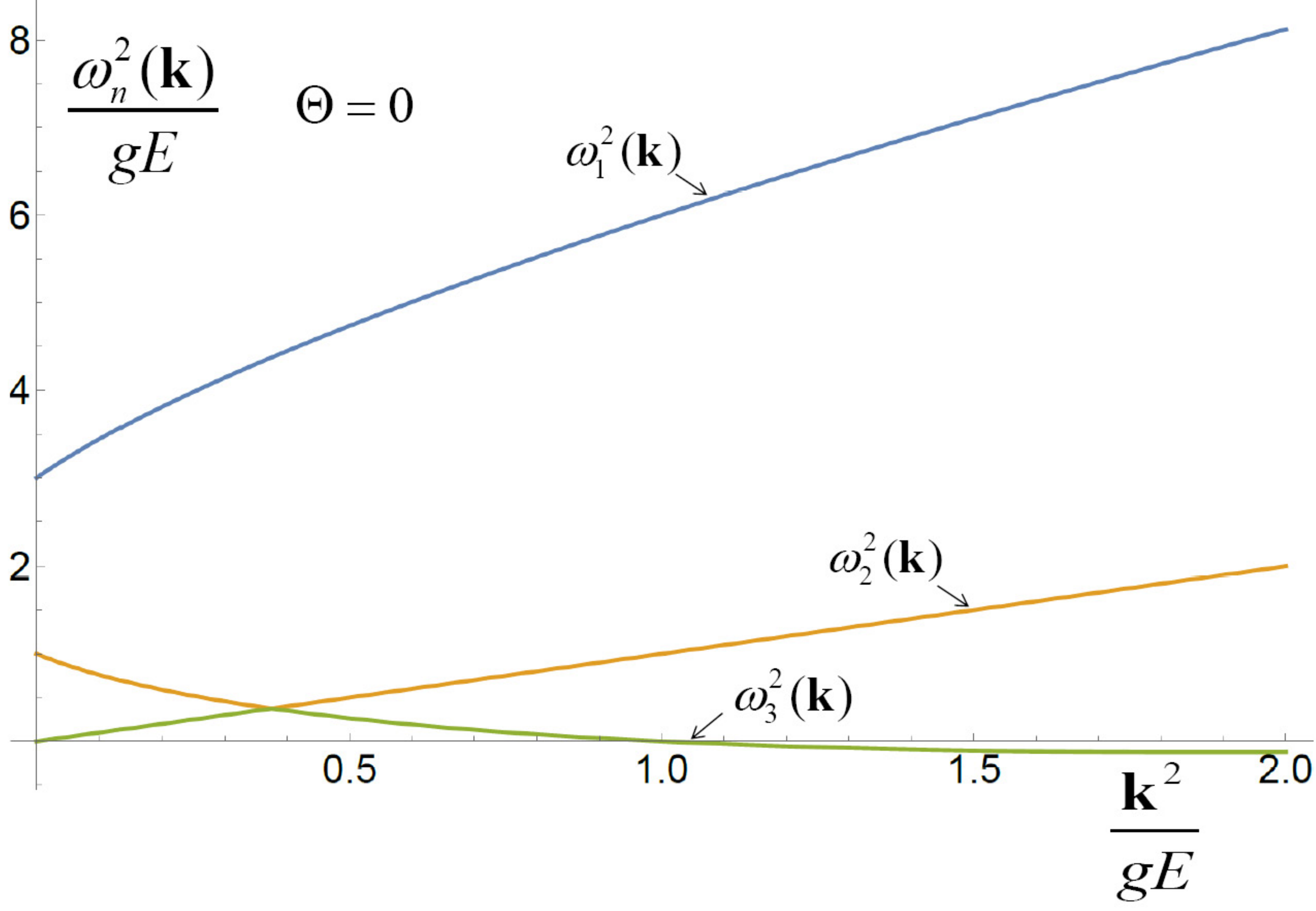}
\vspace{-5mm}
\caption{$\omega_n^2({\bf k})$ as a function of ${\bf k}^2$ for $\Theta = 0$.}
\label{FigE-y-1}
\end{minipage}
\hspace{3mm}
\begin{minipage}{87mm}
\centering
\includegraphics[scale=0.26]{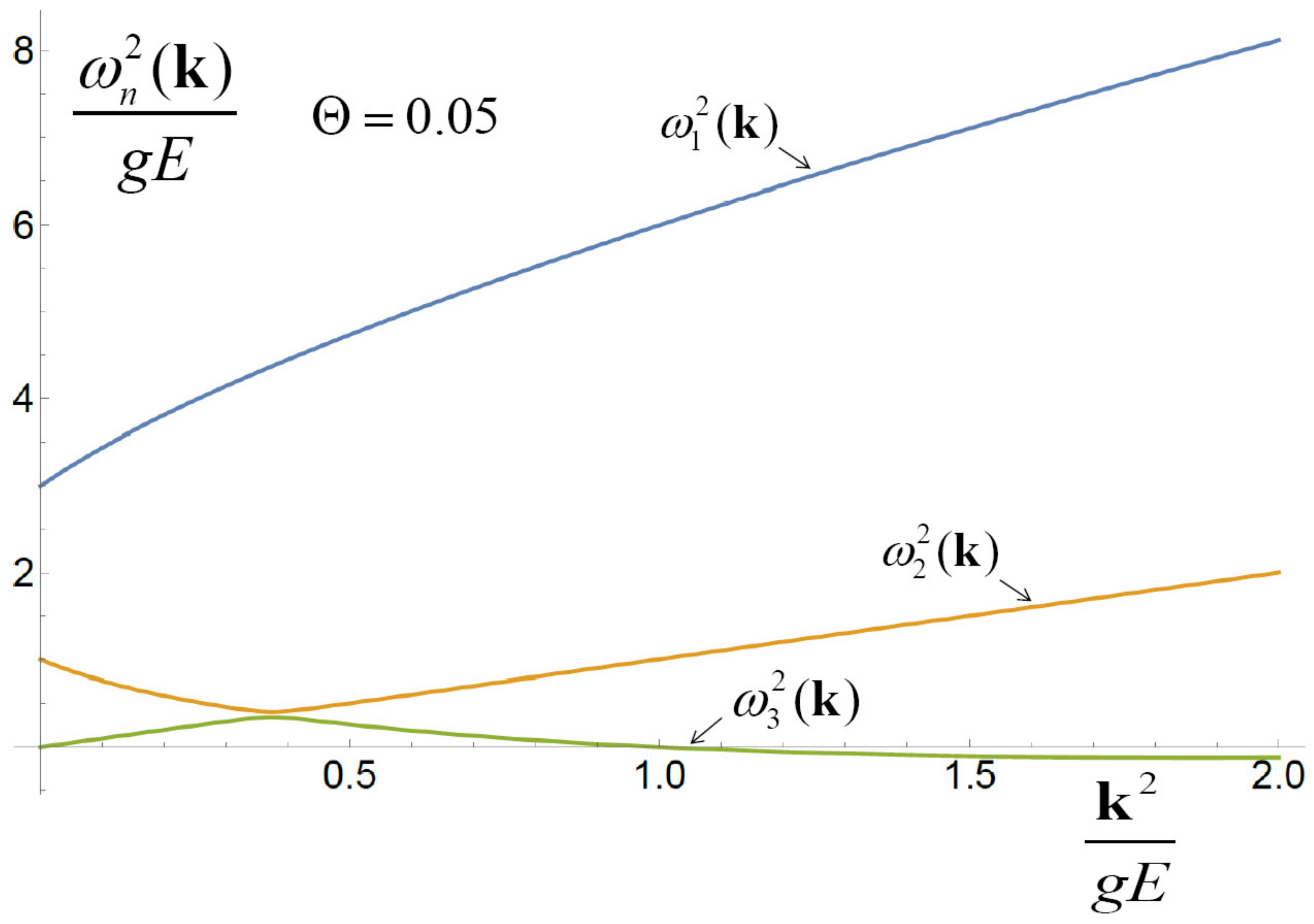}
\vspace{-5mm}
\caption{$\omega_n^2({\bf k})$ as a function of ${\bf k}^2$ for $\Theta = 0.05$.}
\label{FigE-y-2}
\end{minipage}
\end{figure}

The equation of motion of the small field $a^\mu_a$ (\ref{linear-YM-background}) is found to be
\be
\label{a-mu-nA}
\Box a^\mu_a 
+ 2g A (\epsilon^{a2b}\partial_0 + \epsilon^{a3b}\partial_x ) a^\mu_b
+ g^2 A^2 (\epsilon^{a2e}\epsilon^{e2b} - \epsilon^{a3e}\epsilon^{e3b} ) a^\mu_b
+ 2g^2 A^2 \epsilon^{a1b}( \delta^{\mu 0} a^x_b +  \delta^{\mu x} a^0_b) = 0 .
\ee
where $A \equiv \sqrt{E/g}$.

Assuming that $a_a^\mu (t,x,y,z) = e^{-i (\omega t - {\bf k}\cdot {\bf r})} a_a^\mu$, where ${\bf k} = (k_x, k_y, k_z)$ and  ${\bf r} = (x,y,z)$, Eqs.~(\ref{a-mu-nA}) are changed into the following set of algebraic equations
\ba
\label{eq-Etx}
\hat{M}_E^{tx} \,\vec{a^{tx}} &=& 0,
\\
\label{eq-Ey}
\hat{M}_E^y \,\vec{a^y} &=& 0,
\\
\label{eq-Ez}
\hat{M}_E^z \,\vec{a^z} &=& 0,
\ea
where 
\ba
\label{matrix-MEtx}
\hat{M}_E^{tx} =
\left[ {\begin{array}{cccccc}
-\omega^2 + {\bf k}^2 & - 2ig A k_x & -2ig A \omega  & 0 & 0 & 0 \\[2mm]
2ig A k_x & -\omega^2 + {\bf k}^2  + g^2 A^2 & 0 & 0 & 0 & -  2 g^2 A^2 \\[2mm]
2ig A \omega & 0 & -\omega^2 + {\bf k}^2 - g^2 A^2 & 0 & 2 g^2 A^2 & 0 \\[2mm]
0 & 0 & 0 & -\omega^2 + {\bf k}^2  & - 2ig A k_x &  - 2ig A \omega  \\[2mm]
0 & 0 & -2 g^2 A^2 & 2ig A k_x & -\omega^2 + {\bf k}^2 + g^2 A^2 & 0 \\[2mm]
0 & 2 g^2 A^2 & 0 & 2ig A \omega & 0 &  -\omega^2 + {\bf k}^2 - g^2 A^2 \\
 \end{array} } \right]  ,
\ea
\ba
\label{matrix-MEyz}
\hat{M}_E^y = \hat{M}_E^z =
\left[ {\begin{array}{ccc}
-\omega^2 + {\bf k}^2 & - 2ig A k_x & -2ig A \omega   \\[2mm]
2ig A k_x & -\omega^2 + {\bf k}^2  + g^2 A^2 & 0  \\[2mm]
2ig A \omega & 0 & -\omega^2 + {\bf k}^2 -  g^2 A^2 \\
\end{array} } \right]  
\ea
and 
\ba
\vec{a^{tx}} =
\left[ {\begin{array}{c}
a^0_1 \\[2mm]
a^0_2 \\[2mm]
a^0_3 \\[2mm]
a^x_1 \\[2mm]
a^x_2 \\[2mm]
a^x_3 \\
 \end{array} } \right] , 
~~~~~~~~~~~~
\vec{a^y} =
 \left[ {\begin{array}{c}
a^y_1 \\[2mm]
a^y_2 \\[2mm]
a^y_3 \\
 \end{array} } \right]  ,
~~~~~~~~~~~~
\vec{a^z} =
\left[ {\begin{array}{c}
a^z_1 \\[2mm]
a^z_2 \\[2mm]
a^z_3 \\
 \end{array} } \right] .
\ea

Since the equations (\ref{eq-Etx}),  (\ref{eq-Ey}) and  (\ref{eq-Ez}) are all homogeneous, they have solutions if
 \be
{\rm det}\hat{M}_E^{tx}=0, ~~~~~~~~~~
{\rm det}\hat{M}_E^y=0, ~~~~~~~~~~
{\rm det}\hat{M}_E^z=0 ,
\ee
which are the dispersion equations. 

\begin{figure}[t]
\begin{minipage}{87mm}
\centering
\vspace{3mm}
\includegraphics[scale=0.275]{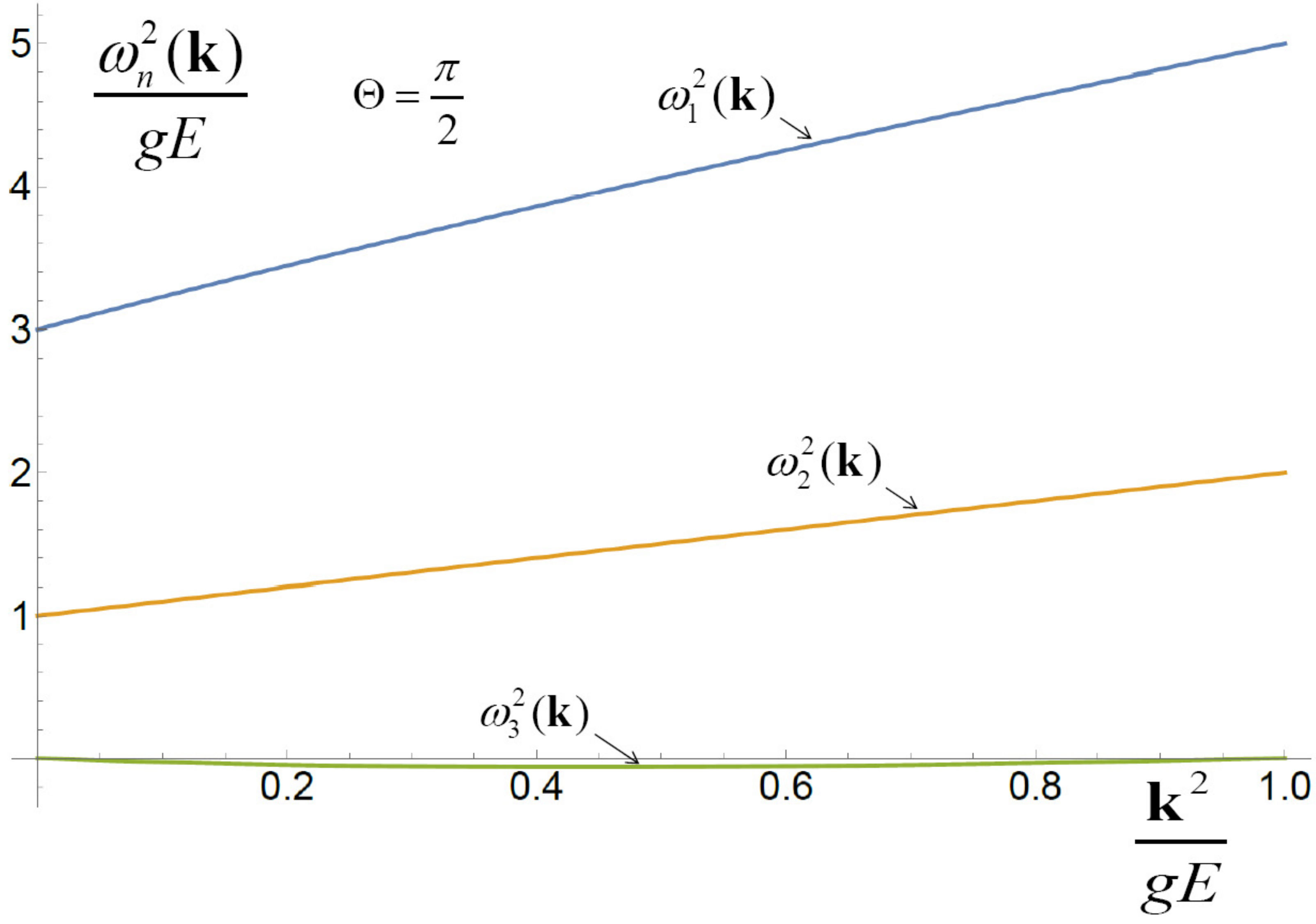}
\vspace{-7mm}
\caption{$\omega_n^2({\bf k})$ as a function of ${\bf k}^2$ for $\Theta = \pi/2$.}
\label{FigE-y-4}
\end{minipage}
\hspace{3mm}
\begin{minipage}{87mm}
\centering
\vspace{-1mm}
\includegraphics[scale=0.255]{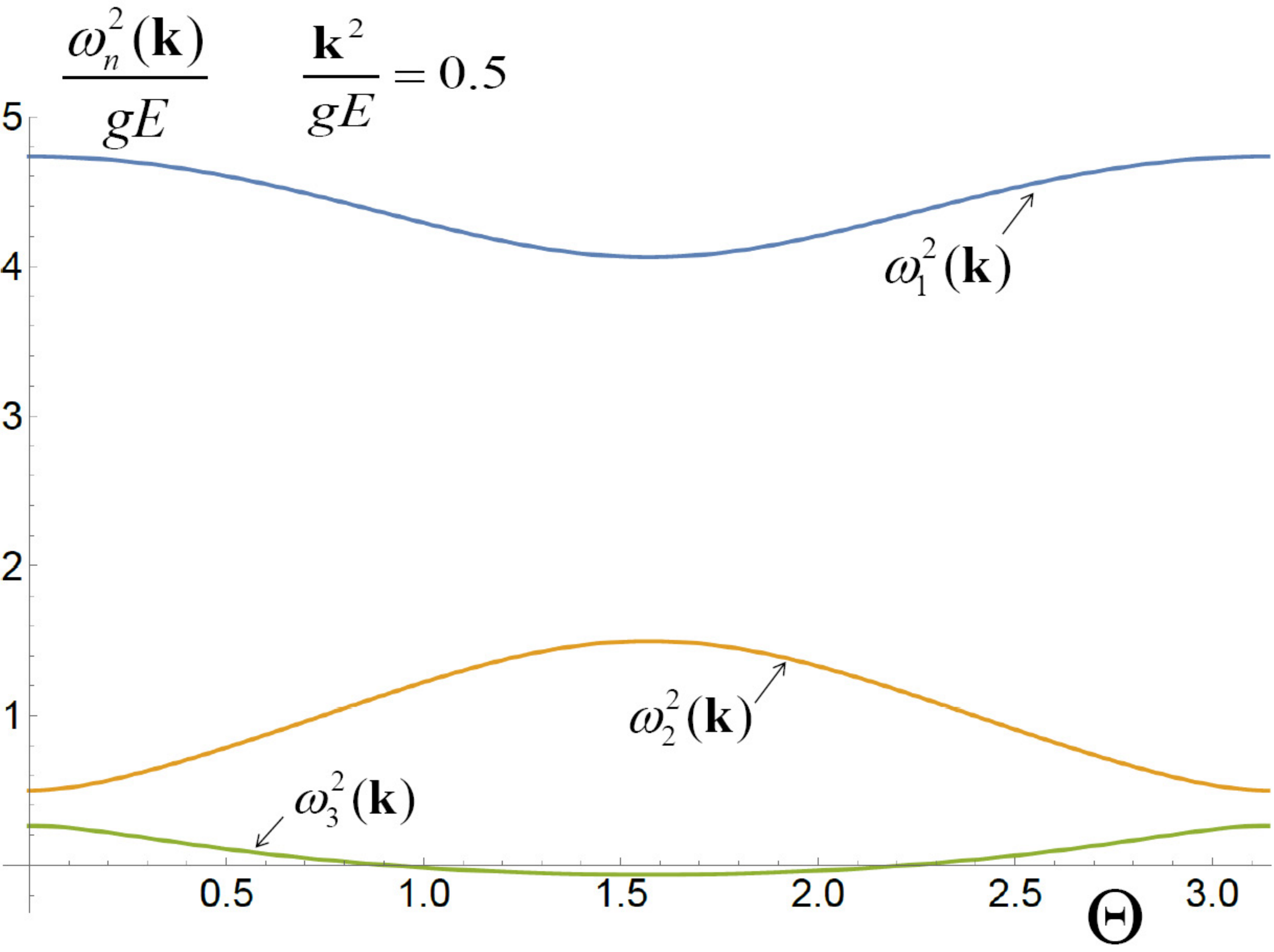}
\vspace{2mm}
\caption{$\omega_n^2({\bf k})$ as a function of $\Theta$ for ${\bf k}^2=0.5 \, gE$.}
\label{FigE-y-5}
\end{minipage}
\end{figure}

\subsection{Equations ${\rm det}\hat{M}_E^y=0$ and ${\rm det}\hat{M}_E^z=0$}
\label{sec-eqs-MEy-MEz}

Let us start with the equation ${\rm det}\hat{M}_E^y=0$. Computing the determinant of the matrix (\ref{matrix-MEyz}) as
\be
{\rm det}\hat{M}_E^y = - \omega^6 
   + (4 g^2 A^2 + 3 k^2 )\omega^4 
   - \big( 3 g^4 A^4  + 4g^2 A^2 (k^2 - k_x^2) + 3k^4\big) \omega^2 
   +k^6 - g^4 A^4 k^2 + 4 g^4 A^4 k_x^2 - 4 g^2 A^2 k^2 k_x^2 ,
\ee
the dispersion equation ${\rm det}\hat{M}_E^y = 0$ is again the cubic equation (\ref{cubic-eq}) but the coefficients are
\ba
\label{a2-Ey}
a_2 &\equiv& - 4 g^2 A^2 - 3 k^2,
\\[2mm]
\label{a1-Ey}
a_1 &\equiv& 3 g^4 A^4  + 4g^2 A^2 (k^2 - k_x^2) + 3 k^4,
\\[2mm]
\label{a0-Ey}
a_0 &\equiv& - k^6 + g^4 A^4 k^2 - 4 g^4 A^4 k_x^2 + 4 g^2 A^2 k^2 k_x^2 .
\ea

The discriminant equals
\ba
\label{discriminant-final}
\frac{1}{4g^6 A^6} \, \Delta = 9 g^6 A^6 
+ 4 g^4 A^4 (13 k^2 - 7 k_x^2)  
+ 4 g^2 A^2 (25 k^4 + 14 k^2 k_x^2 - 119 k_x^4) 
+ 64 (k^2 +k_x^2)^3,
\ea
and it is positive. Consequently, the solutions, which are real, can be written, as previously, in the Vi\` ete's trigonometric form (\ref{x-n}).

\begin{figure}[t]
\begin{minipage}{87mm}
\centering
\includegraphics[scale=0.36]{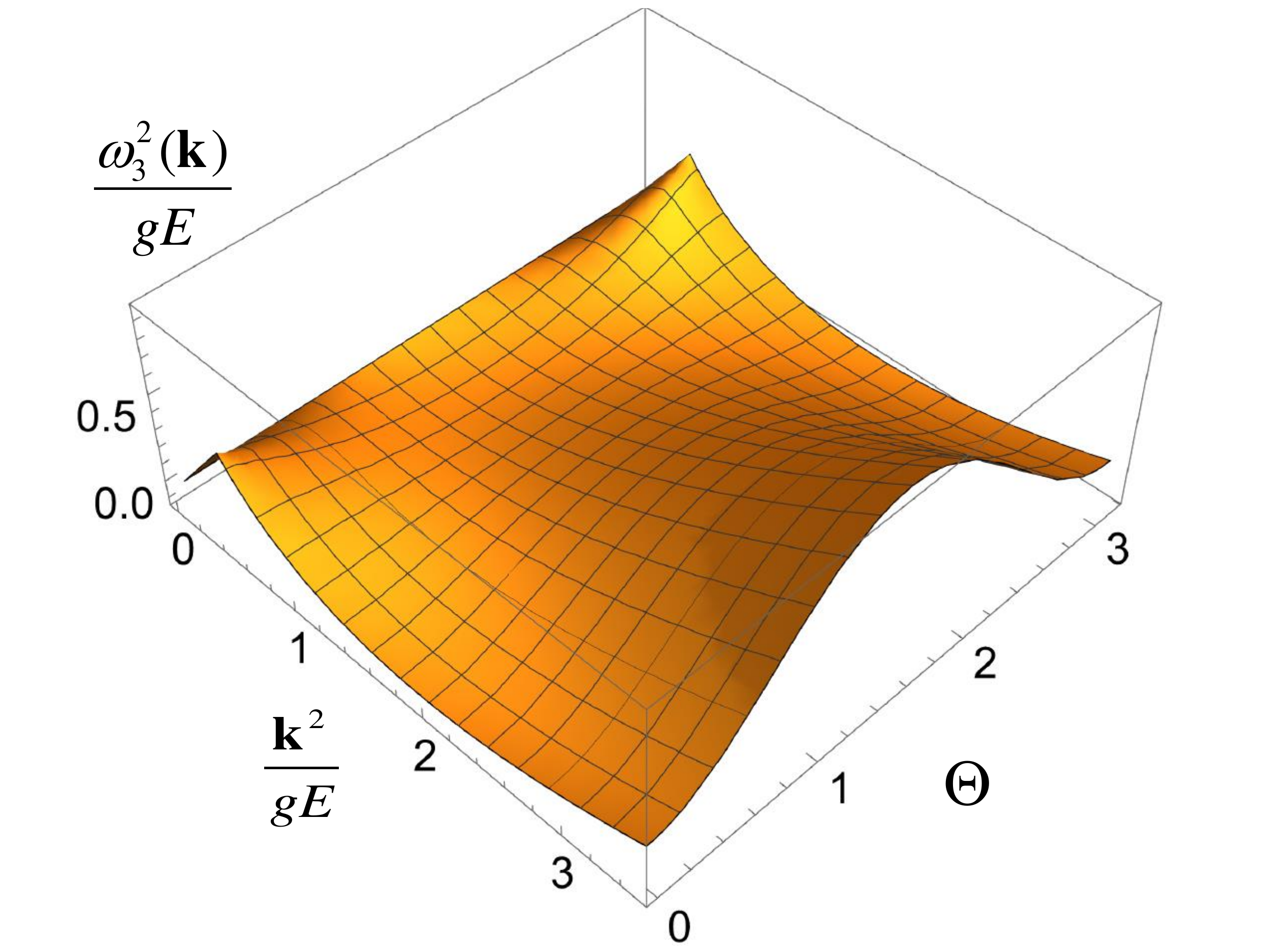}
\end{minipage}
\hspace{2mm}
\begin{minipage}{87mm}
\centering
\includegraphics[scale=0.35]{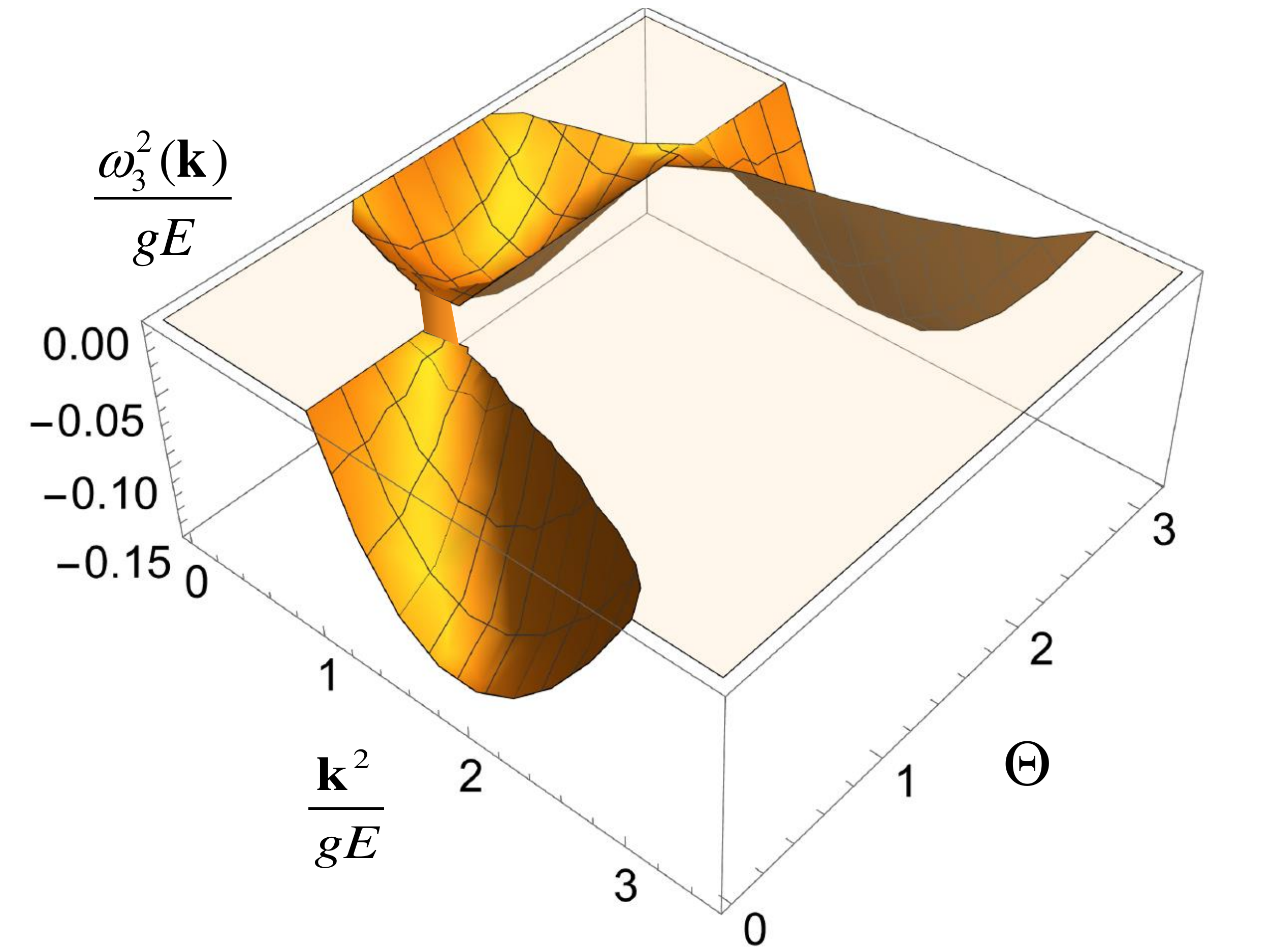}
\end{minipage}
\vspace{-1mm}
\caption{$\omega_3^2({\bf k})$ (left panel) and negative part of $\omega_3^2({\bf k})$ (right panel) as functions of $\Theta$ and ${\bf k}^2$.}
\label{FigE-y-7}
\end{figure}

The solutions are shown in Figs.~\ref{FigE-y-1} - \ref{FigE-y-7}. The $x$ component of the wave vector ${\bf k}$ is expressed as $k_x = k \cos\Theta$ and the solutions are shown as functions of ${\bf k}^2$ or $\Theta$. The spectrum of modes of the equation ${\rm det}\hat{M}_E^y = 0$ is rather complex. In Figs.~\ref{FigE-y-1} and \ref{FigE-y-2} one observes the {\it mode coupling} of $\omega_2^2({\bf k})$ and $\omega_3^2({\bf k})$. One could think that the curves of $\omega_2^2({\bf k})$ and $\omega_3^2({\bf k})$ computed for $\Theta=0$ and shown in Fig.~\ref{FigE-y-1} cross each other. However, when the curves are computed for $\Theta=0.05$ and shown in Fig.~\ref{FigE-y-2} one sees that the curves instead only approach each other. For bigger values of $\Theta$ the curves $\omega_2^2({\bf k})$ and $\omega_3^2({\bf k})$ are well separated. The phenomenon of mode coupling is discussed in detail and explained in \S 64  of  the textbook \cite{Landau-Lifshitz-1981}. 

One observes in Figs.~\ref{FigE-y-1} - \ref{FigE-y-5} that $\omega_3^2({\bf k})$ can be negative. Then, there is a pair of pure imaginary modes, one is unstable and one is overdamped. We show $\omega_3^2({\bf k})$ as a function of ${\bf k}^2$ and $\Theta$ in the left panel of Fig.~\ref{FigE-y-7}. In the right panel of Fig.~\ref{FigE-y-7} one sees the domain of ${\bf k}^2$ and $\Theta$ where $\omega_3^2({\bf k})$ is negative. The solutions of the equation ${\rm det}\hat{M}_E^z=0$ are obviously the same as those of ${\rm det}\hat{M}_E^y=0$.

\subsection{Equation ${\rm det}\hat{M}_E^{tx}=0$}

Let us now discuss the equation ${\rm det}\hat{M}_E^{tx}=0$. We start with the simple special ${\bf k}=0$ when the determinant of the matrix (\ref{matrix-MEtx}) is computed as
\be
{\rm det}\hat{M}_E^{tx} = \omega^4 (\omega^4 - 4 g^2 A^2 \omega^2 + 7 g^4 A^4 )^2.
\ee
The solutions of the equation ${\rm det}\hat{M}_E^{tx}=0$ are: the double solution $\omega^2=0$ and double solutions
\be
\label{sol-a0x-k=0-quadratic}
\omega^2_{\pm} = (2 \pm i \sqrt{3}\,) g^2A^2,
\ee
which give the mode frequencies
\be
\label{sol-a0x-k=0}
\omega_{(+,\pm )} = \pm 7^{1/4} gA \big(\cos(\phi/2) + i \sin(\phi/2) \big),
~~~~~~~~~~~~
\omega_{(-,\pm )} = \pm 7^{1/4} gA \big(\cos(\phi/2) - i \sin(\phi/2) \big),
\ee
where
\be
\phi = {\rm arctg}\Big(\frac{\sqrt{3}}{2}\Big) .
\ee
One observes that the modes $\omega_{(+,+)}$ and $\omega_{(-,-)}$ are unstable as their imaginary parts are positive. 

\begin{figure}[t]
\begin{minipage}{87mm}
\centering
\includegraphics[scale=0.34]{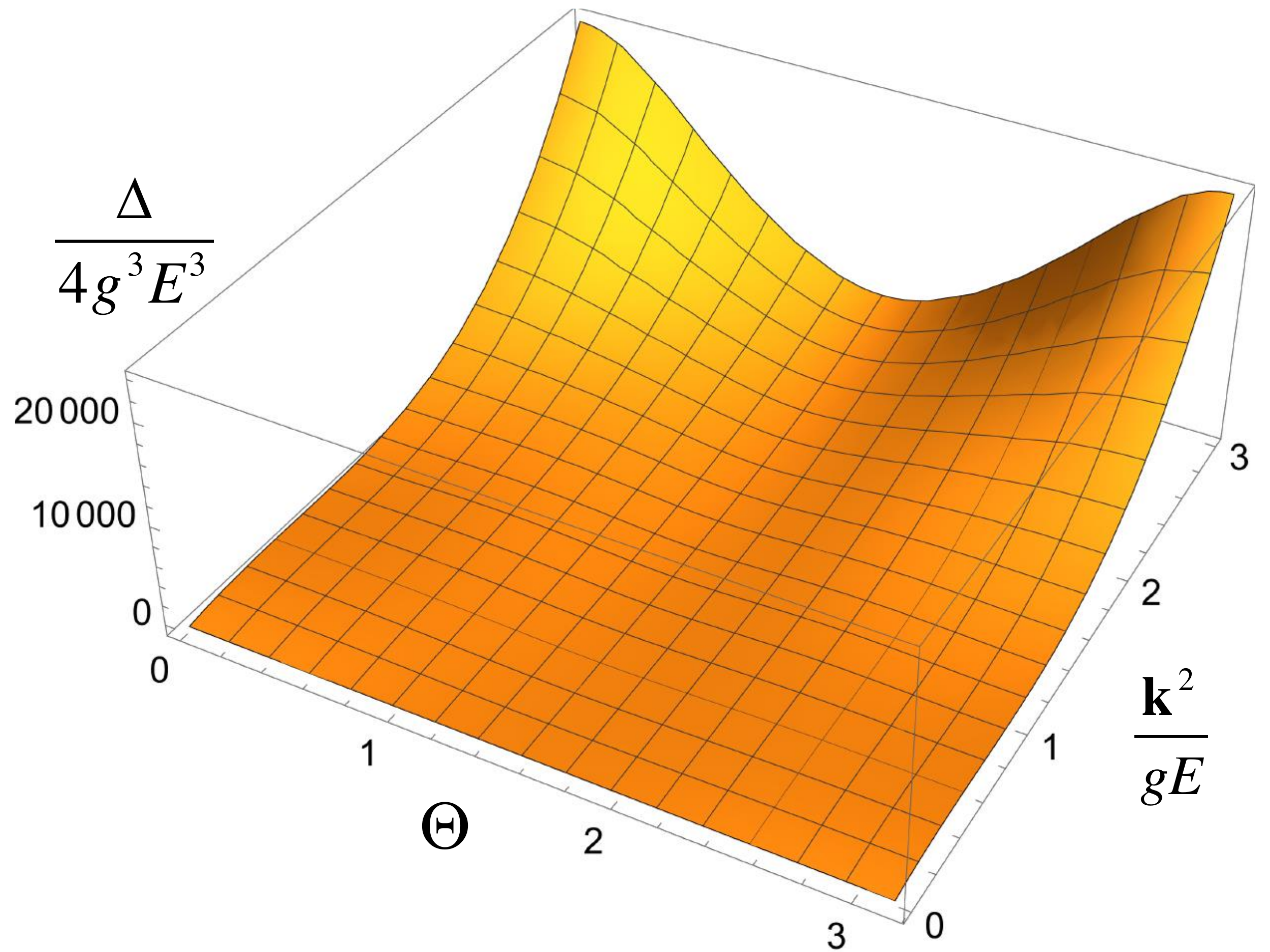}
\vspace{-6mm}
\caption{$\frac{\Delta}{4g^3 E^3}$ as a function of $\Theta$ and ${\bf k}^2$.}
\label{FigE-tx-1}
\end{minipage}
\hspace{2mm}
\begin{minipage}{87mm}
\centering
\includegraphics[scale=0.34]{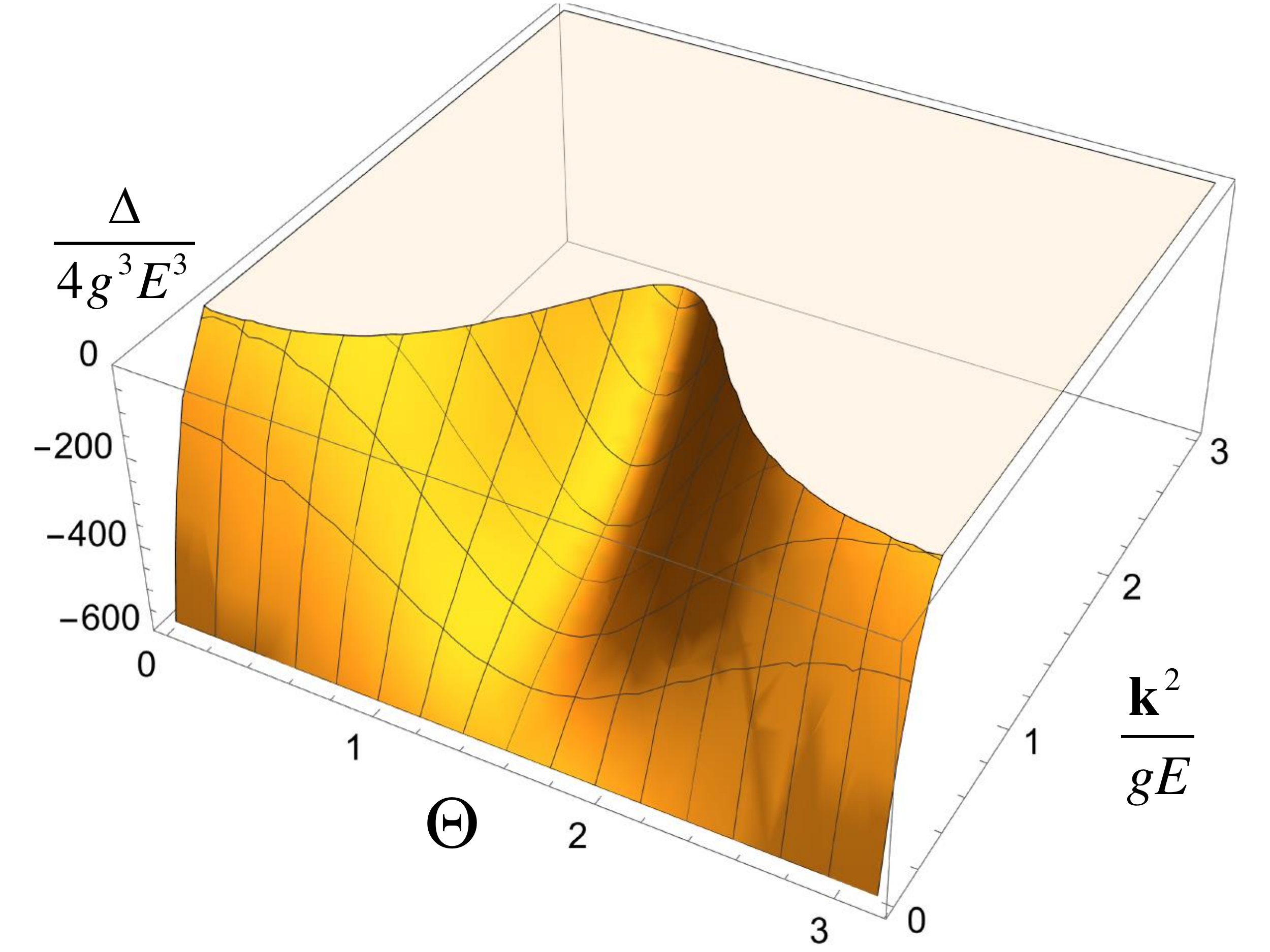}
\vspace{-6mm}
\caption{Negative part of $\frac{\Delta}{4g^3 E^3}$ as a function of $\Theta$ and ${\bf k}^2$.}
\label{FigE-tx-2}
\end{minipage}
\end{figure}

The general case of ${\bf k} \not= 0$ is much more complicated than the  ${\bf k} = 0$ case, but there is an important simplification. The determinant of the matrix (\ref{matrix-MEtx}), which is the polynomial of $\omega^2$ of order 6, appears to be the square of the polynomial of order 3 that is 
\ba
\label{det-MEtx}
&& {\rm det}\hat{M}_E^{tx} 
\\[2mm]\nn
&& = 
\big[- \omega^6 + (4 g^2 A^2 + 3 k^2) \omega^4
 - (7 g^4 A^4 + 4 g^2 A^2 (k^2 - k_x^2) + 3 k^4) \omega^2 
+3 g^4 A^4 k^2 + 4 g^4 A^4 k_x^2 - 4 g^2 A^2 k^2 k_x^2 + k^6 \big]^2.
\ea
So, we have again the cubic equation (\ref{cubic-eq}) in $\omega^2$ with the coefficients 
\ba
\label{a2-E-tx}
a_2 &\equiv& - 4 g^2 A^2 - 3 k^2,
\\[2mm]
\label{a1-E-tx}
a_1 &\equiv& 7 g^4 A^4  + 4g^2 A^2 (k^2 - k_x^2) + 3 k^4,
\\[2mm]
\label{a0-E-tx}
a_0 &\equiv& - 3 g^4 A^4 k^2 - 4 g^4 A^4 k_x^2 + 4 g^2 A^2 k^2 k_x^2 - k^6.
\ea
We note that because of the square in the determinant (\ref{det-MEtx}) each solution of the cubic equation is doubled. 

The discriminant $\Delta$ with the coefficients (\ref{a2-E-tx}), (\ref{a1-E-tx}) and (\ref{a0-E-tx}) equals
\be
\frac{\Delta}{4g^6 A^6} = - 147 g^6 A^6 + 4 g^4 A^4 (29 k^2 + 153 k_x^2)
- 4 g^2 A^2 (23 k^4 + 82 k^2 k_x^2 + 167 k_x^4) + 64 (k^2 + k_x^2)^3.
\ee
It can be either positive or negative as shown in Figs.~\ref{FigE-tx-1} and \ref{FigE-tx-2}. 

\begin{figure}[t]
\begin{minipage}{87mm}
\centering
\includegraphics[scale=0.27]{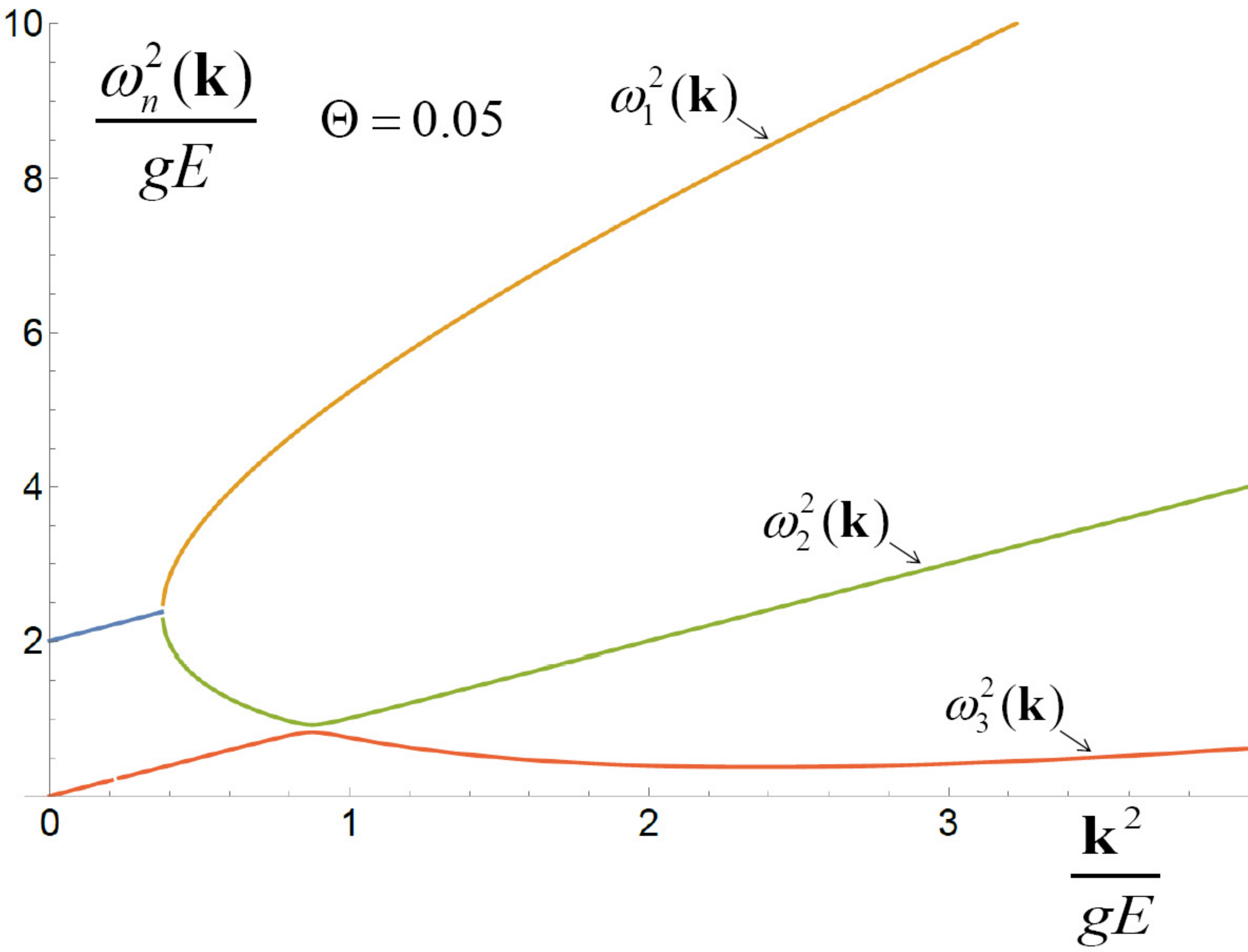}
\vspace{-2mm}
\caption{$\omega_n^2({\bf k})$ as a function of ${\bf k}^2$ for $\Theta = 0.05$.}
\label{FigE-tx-3}
\end{minipage}
\hspace{2mm}
\begin{minipage}{87mm}
\centering
\vspace{-1mm}
\includegraphics[scale=0.27]{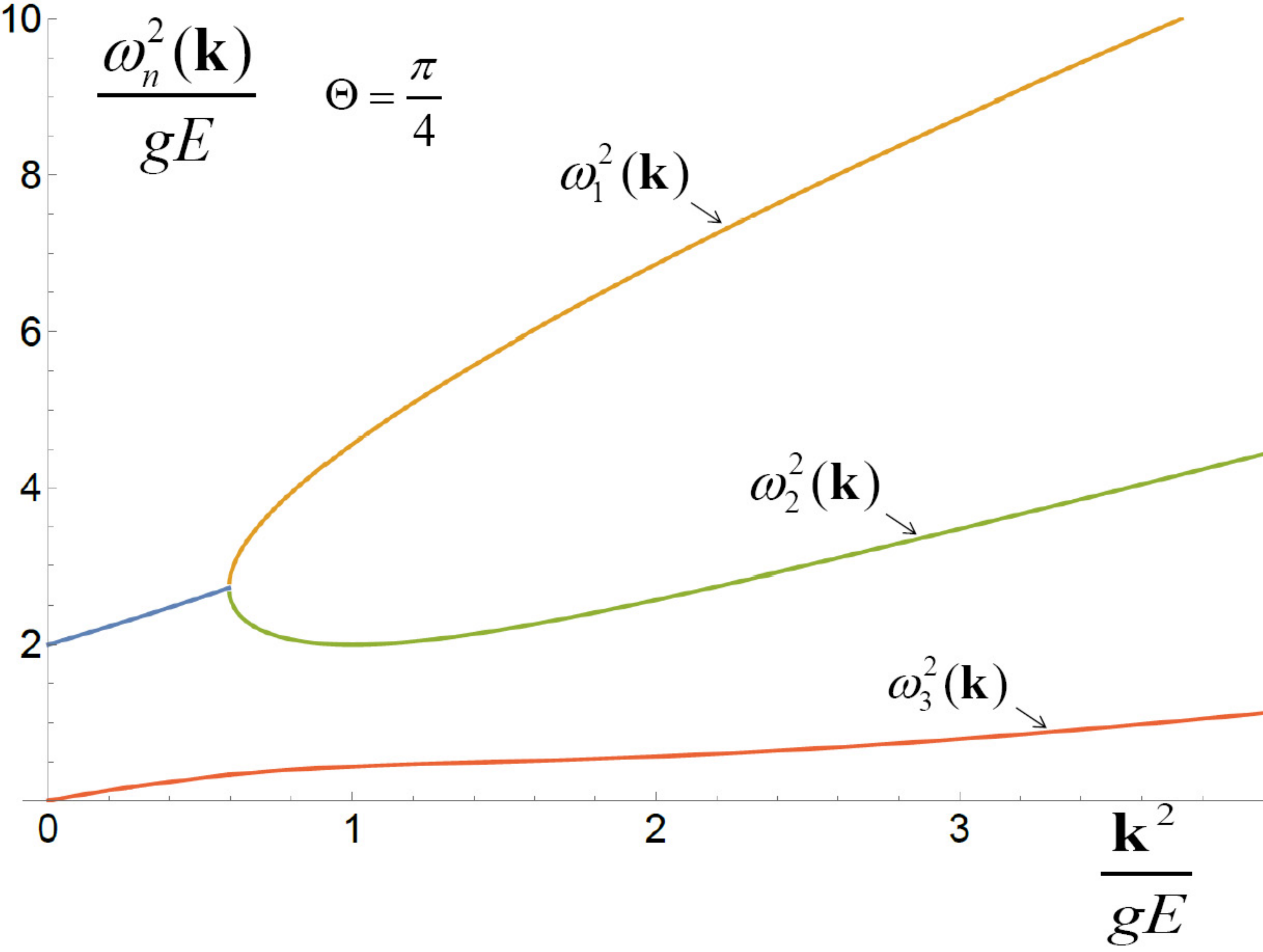}
\vspace{-6mm}
\caption{$\omega_n^2({\bf k})$ as a function of ${\bf k}^2$ for $\Theta = \pi/4$.}
\label{FigE-tx-4}
\end{minipage}
\end{figure}

When $\Delta > 0$ the solutions of the cubic equation are real and can be expressed in the Vi\` ete's trigonometric form (\ref{x-n}). When $\Delta < 0$ there are one real and two complex solutions which can be found using the Cardano formula \cite{Bronshtein-Semendyayev-1985}. The solutions are
\ba
x_1 &=& -\frac{1}{2}(u + v) + \frac{i \sqrt{3}}{2}(u - v) - \frac{1}{3}a_2,
\\
x_2 &=& -\frac{1}{2}(u + v) - \frac{i \sqrt{3}}{2}(u - v) - \frac{1}{3}a_2,
\\
x_3 &=& u + v  - \frac{1}{3}a_2,
\ea
where 
\be
u \equiv \sqrt[3]{-\frac{q}{2} + \sqrt{\frac{q^2}{4} + \frac{p^3}{27}}} ,
~~~~~~~~~~
v \equiv \sqrt[3]{-\frac{q}{2} - \sqrt{\frac{q^2}{4} + \frac{p^3}{27}}} .
\ee

\begin{figure}[t]
\begin{minipage}{87mm}
\centering
\includegraphics[scale=0.27]{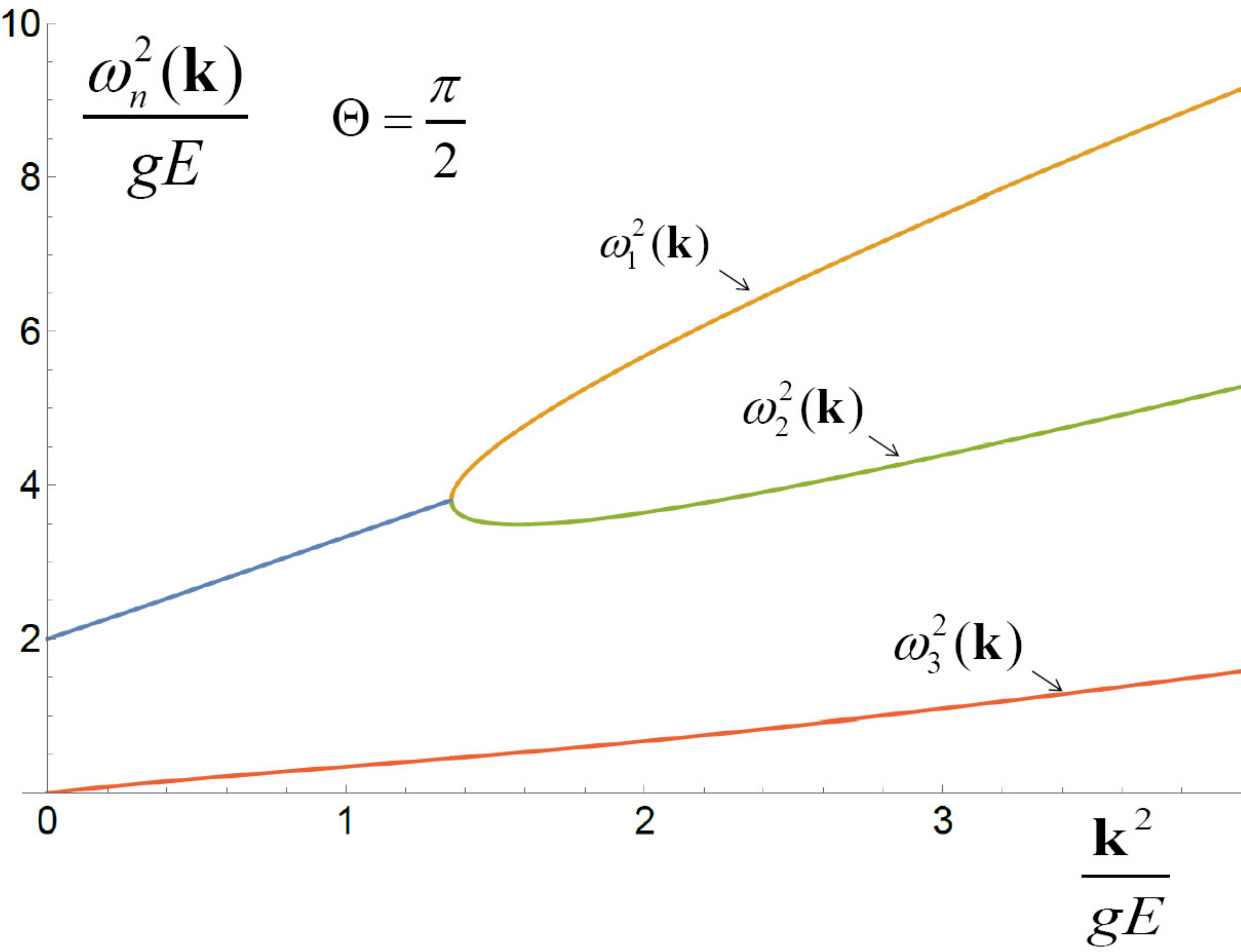}
\vspace{-6mm}
\caption{$\omega_n^2({\bf k})$ as a function of ${\bf k}^2$ for $\Theta = \pi/2$.}
\label{FigE-tx-5}
\end{minipage}
\hspace{2mm}
\begin{minipage}{87mm}
\centering
\vspace{5mm}
\includegraphics[scale=0.27]{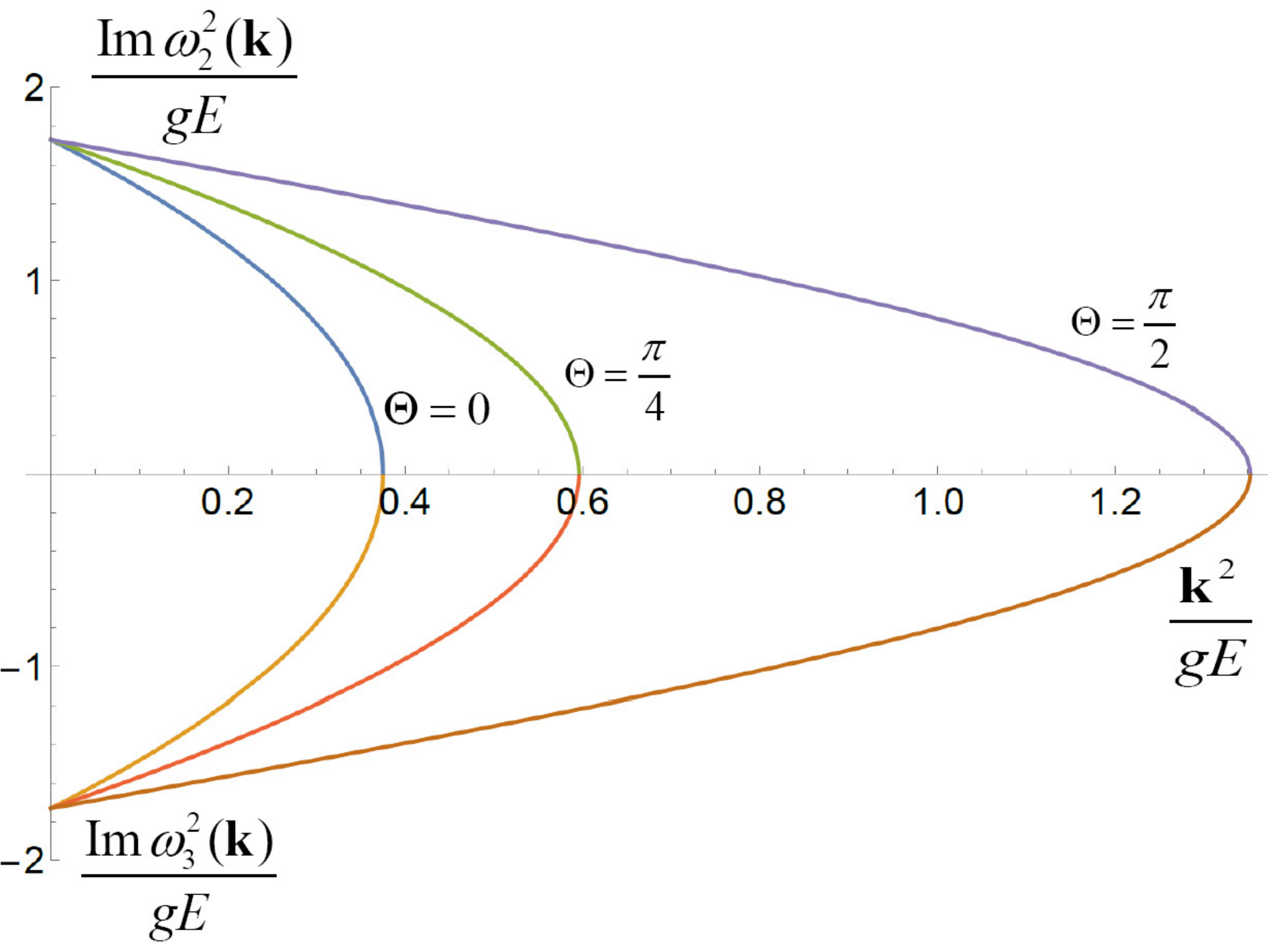}
\vspace{-7mm}
\caption{Imaginary parts of $\omega_1^2({\bf k})$ and $\omega_2^2({\bf k})$ as functions of ${\bf k}^2$ for $\Theta = 0,\,\pi/4, \,\pi/2$.}
\label{FigE-tx-6}
\end{minipage}
\hspace{2mm}
\end{figure}

The solutions in both regions $\Delta < 0$ and $\Delta > 0$ are shown in Figs.~\ref{FigE-tx-3}-\ref{FigE-tx-6}. Figs.~\ref{FigE-tx-3}-\ref{FigE-tx-5} present the real solutions from the domain of $\Delta > 0$ combined with the real parts of the complex solutions from $\Delta < 0$. The solutions are shown as functions of ${\bf k}^2$ for three values of $\Theta = 0.05, \, \pi/4, \, \pi/2$. In  Fig.~\ref{FigE-tx-6} we show the imaginary parts of the complex solutions from the domain  $\Delta < 0$. The whole spectrum of the solutions of the equation ${\rm det}\hat{M}_E^{tx}=0$ is far not simple. 

We first note that the real parts of  $\omega_1^2({\bf k})$ and $\omega_2^2({\bf k})$ are equal to each other (${\rm Re}\omega_1^2({\bf k}) = {\rm Re}\omega_2^2({\bf k})$) in the region $\Delta < 0$ while the imaginary parts shown in Fig.~\ref{FigE-tx-6} are of the opposite sign (${\rm Im}\omega_1^2({\bf k}) =- {\rm Im}\omega_2^2({\bf k})$).

One sees in Figs.~\ref{FigE-tx-3}-\ref{FigE-tx-5} that the real part of low momentum solution $\omega_1^2({\bf k})$ or $\omega_2^2({\bf k})$ from the region $\Delta < 0$ bifurcates into two real solutions $\omega_1^2({\bf k})$ and $\omega_2^2({\bf k})$ from the region $\Delta > 0$. The bifurcation occurs at $\Delta = 0$.

The solution $\omega_3^2({\bf k})$ is real in both regions $\Delta < 0$ and $\Delta > 0$ and it is smooth at $\Delta = 0$, as seen in Figs.~\ref{FigE-tx-3}-\ref{FigE-tx-5}. Fig.~\ref{FigE-tx-3} shows that, as in case of the equation ${\rm det}\hat{M}_E^y=0$ discussed in Sec.~\ref{sec-eqs-MEy-MEz}, the solution $\omega_2^2({\bf k})$ is coupled to $\omega_3^2({\bf k})$.

There are two unstable modes which occur as square roots of $\omega_1^2({\bf k})$ and $\omega_2^2({\bf k})$ from the region $\Delta < 0$. These are the modes with positive imaginary parts which at ${\bf k}=0$ are equal to $\omega_{(+,+)}$ and $\omega_{(-,-)}$ from Eq.~(\ref{sol-a0x-k=0}). Fig.~\ref{FigE-tx-6} shows that the imaginary parts are maximal at ${\bf k}=0$ and equal to ${\rm  Im}\omega_{(+,+)} = {\rm  Im}\omega_{(-,-)}$. So, the fastest modes grow as $e^{\gamma t}$ with $\gamma = {\rm Im} \omega_{(+,+)} \approx 0.57 \sqrt{gE}$.

\section{Summary, Discussion and Outlook}
\label{sec-final}

Linear stability of classical magnetic and electric fields, which are constant and uniform, have been studied. We have considered the Abelian configurations where the fields are generated by a single-color potential linearly depending on coordinates and the nonAbelian configurations where the fields occur due to constant non-commuting potentials of different colors. While the potentials of Abelian configurations solve the sourceless Yang-Mills equations the nonAbelian configurations require  non-vanishing currents to satisfy the equations. All four cases have been analyzed using the same background gauge condition. The Abelian and nonAbelian configurations are physically inequivalent and indeed the spectra of eigenmodes of small fluctuations around the background fields are different though similar. The spectra include unstable modes in all cases. 

One asks how our analysis changes when the fields are uniform not in the infinite coordinate space but only in a limited domain of a size $L$? Assuming that the domain is a cube centered at ${\bf r}=0$ and demanding that the real potentials $a_a^\mu$ vanish at the edge of the cube, the wave vectors $k_x, k_y, k_z$ should be replaced as
\be
\label{replacement}
(k_x,k_y,k_z) \longrightarrow (2l_x +1, 2l_y +1, 2l_z +1) \frac{\pi}{L} , 
\ee
where $l_x, l_y,l_z$ are integer numbers. Consequently, a spectrum of eigenmodes becomes discrete and some unstable modes can disappear. As an illustrative example we consider the unstable mode of the Abelian configuration of magnetic field which is $\omega^2 = - gB + k_x^2$.  After the replacement (\ref{replacement}) it becomes $\omega^2 = - gB + \pi^2 (2l_x+1)^2/L^2$. One sees that $\omega^2 < 0$ at least for $l_x=0$ if $gB L^2 > \pi^2$. So, the instability exists if the field is sufficiently strong and uniform over a sufficiently big domain. Since all unstable modes we found occur at small wave vectors there are analogous conditions for existence of instabilities of the fields which are uniform only in a limited spatial domain. 

We intend to extend our analysis to the case of parallel electric and magnetic fields which are both present at the same time. Such a situation is expected at the earliest phase of relativistic heavy-ion collisions described in the framework of Color Glass Condensate. To make our considerations more relevant for relativistic heavy-ion collisions we plan to analyze stability not of the fields which are constant and uniform but rather the fields which are invariant under Lorentz boosts along the collision axis.  

\section*{Acknowledgments}

This work was partially supported by the National Science Centre, Poland under grant 2018/29/B/ST2/00646. 


\end{document}